\documentclass[11pt,3p]{elsarticle}
\usepackage{verbatim}
\usepackage{lineno,hyperref}
\modulolinenumbers[5]
\usepackage{color}
\journal{CARBON (Elsevier)}
\makeatletter
\def\ps@pprintTitle{%
 \let\@oddhead\@empty
 \let\@evenhead\@empty
 \def\@oddfoot{}%
 \let\@evenfoot\@oddfoot}
\makeatother
\usepackage{siunitx}
\usepackage[]{subfigure}
\usepackage{placeins}
\usepackage[english]{babel}
\usepackage[utf8x]{inputenc}
\usepackage{graphicx}
\usepackage{booktabs}
\usepackage{tabularx}
\usepackage{caption}
\usepackage{amsmath}
\usepackage{mathtools}
\usepackage{float}
\usepackage{xcolor}
\usepackage{telprint}
\usepackage{natbib}

\begin{document}

\begin{frontmatter}

\title{Monte Carlo simulations of measured electron energy-loss spectra of diamond and graphite: role of dielectric-response models.}

\author[TIPFA,DICAM]{Martina Azzolini}
\author[TIPFA,DICAM]{Tommaso Morresi}

\author[TIPFA]{Giovanni Garberoglio}
\author[TIPFA]{Lucia Calliari}

\author[DICAM,LONDON,BIO]{Nicola M. Pugno}

\author[TIPFA,PRA]{Simone Taioli\corref{mycorrespondingauthor}}
\cortext[mycorrespondingauthor]{Corresponding authors}
\ead{taioli@ectstar.eu}
\author[TIPFA]{Maurizio Dapor\corref{mycorrespondingauthor}}
\ead{dapor@ectstar.eu}
\address[TIPFA]{European Centre for Theoretical Studies in Nuclear Physics and Related Areas (ECT*-FBK) and Trento Institute for Fundamental Physics and Applications (TIFPA-INFN), Trento, Italy}
\address[DICAM]{Laboratory of Bio-inspired \& Graphene Nanomechanics, Department of Civil, Environmental and Mechanical Engineering, University of Trento, Italy}
\address[PRA]{Faculty of Mathematics and Physics, Charles University, Prague, Czech Republic}
\address[LONDON]{School of Engineering and Materials Science, Queen Mary University of London, UK}
\address[BIO]{Centre for Materials and Microsystems, Fondazione Bruno Kessler, Trento, Italy}

\begin{abstract}
In this work we compare Monte Carlo (MC) simulations of electron transport properties with reflection electron energy loss  measurements in diamond and graphite films. 
We assess the impact of different approximations of the dielectric response on the observables of interest for the characterization of carbon-based materials.
We calculate the frequency-dependent dielectric response and energy loss function of these materials in two ways: a full ab initio approach, in which we carry out time-dependent density functional simulations in linear response for different momentum transfers, and a semi-classical model, based on the Drude--Lorentz extension to finite momenta of the optical dielectric function. 
Ab initio calculated dielectric functions lead to a better agreement with electron energy loss measured spectra with respect the widely used Drude-Lorentz model. 
This discrepancy is particularly evident for insulators and semiconductors beyond the optical limit ($\mathbf{q} \neq 0$), where single particle excitations become relevant. Furthermore, we show that the behaviour of the energy loss function at different accuracy levels has a dramatic effect on other physical observables, such as the inelastic mean free path and the stopping power in the low energy regime ($< 100$ eV) and thus on the accuracy of MC simulations.
\end{abstract}
 
\end{frontmatter}

\section{Introduction}
Electron beams are widely used in the characterization of materials to probe composition and structure as well as in the fabrication of electronic devices via plasma-based etching processes and lithography. 
In particular, the success of electron microscopy and electron spectroscopy in reaching high resolving power is due on the one hand to the much smaller electron beam wavelength in comparison to light;
and on the other hand to the easiness of handling electrons, which can be detected, counted and analyzed with respect to energy and angular distribution using electromagnetic fields.  
Therefore, the study of the electron transport inside materials is of paramount importance to understand and to control the interaction mechanisms and energy-transfer scattering processes that occur at different energy ranges in electronic devices \citep{Dapor_FMATS_2015,taioli2010electron,taioli2015chapter}.\\
Moreover, the synthesis of novel carbon allotropes, such as nanotubes and graphene, by chemical vapour deposition (CVD) on metals \citep{taioli2014computational} or by mechanical exfoliation resulted in a renovated interest in carbon-based electronics  \citep{taioli2009electronic,umari2012communication,Phaedon}. Nevertheless, growth techniques, mainly based on catalytic processes on metallic substrates, are still largely debated  \citep{taioli2016characterization}, being of particular concern the graphene flake after-growth transfer to different semiconductor substrates  and the high working temperature of heteroepitaxial synthesis approaches \citep{tatti2016synthesis}. Unfortunately, difficulties encountered in growing high-quality graphene flakes currently hamper the theoretical potential of this 2D material. Furthermore, the use of graphene in micro-electronic applications forces the opening of a band gap \citep{haberer2010tunable,haberer2011direct}.\\
However, other stable or naturally occurring allotropic forms of carbon, such as diamond, multi-layer graphene, and graphite, along with its intercalation compounds, could be used as viable candidates for an all-carbon electronics revolution.
In particular, diamond was long considered an ideal candidate for enhancing the performances of electronic devices due to its high thermal conductivity and charge mobility, wide band gap, optical isotropic structure and robustness owing to its strong covalent $sp^3$-hybridized structure.
On the other hand, graphite is the most stable carbon allotrope, arranged in the form of a layered solid, showing both strong two-dimensional $sp^2$-hybridized lattice bonds, similar in strength to those found in diamond, and weak interplanar bonds that make it soft and malleable as well as anisotropic to external perturbations. Furthermore, graphite shows optimal heat and electricity conductivity retaining the highest natural strength and stiffness even at temperatures in excess of 3000$^{\circ}$ C. \\ 
In this respect, this work is aimed at modelling the electron transport properties of diamond and graphite films by calculating a number of observables of paramount importance for designing novel optical and electronic devices, such as inelastic mean free path, stopping power, plasmons and secondary electron spectra. The specific goal of our analysis is to unravel the impact that different theoretical approaches for calculating the dielectric function, ranging from ab-initio calculations to the use of parametrized dielectric function, such as Drude -- Lorentz model (DL), may have on the assessment of the dielectric response of these two materials by comparing our simulations with experimental reflection electron
energy loss spectra (REELS). This study represents thus a fundamental step towards a better understanding of the basic properties characterizing both bulk and thin film carbon materials as well as an important 
contribution towards the development of all-carbon electronics. \\
We notice that all the observables considered in our analysis are based on the accurate assessment of the frequency-dependent dielectric function, 
which provides the link between microscopic properties, such as the band structure of solids, and macroscopic features that are a direct outcome of spectroscopic experiments, 
such as the absorption coefficient, the surface impedance or the electron energy loss.\\
To compute the dielectric function dependence on the transferred momentum we proceed along three different routes: i) first, we use a semi-classical approach, whereby one assumes the knowledge of the long wavelength or optical limit of the dielectric function ($q \rightarrow 0$ limit); this information is usually provided by experimental measurements of optical absorption \citep{PhysRevB.90.205422} or ab initio simulations \citep{Baroni}. To go beyond the optical limit, 
we extend the dielectric response to finite momenta by using a DL model. In this approach the dielectric function is approximated by a number of damped harmonic oscillators with frequencies equal to the plasmon frequencies obtained by fitting experimental data \citep{DaporGarb_MPhyResB,Dapor_NIMB_2007} and a friction-type force to simulate general dissipative processes. ii) the second approach uses ab initio simulations to calculate the dielectric response for vanishing momentum transfer, eventually extended to finite momenta by a DL model; iii) third, we find the dispersion law of the dielectric function at finite momentum $q$ by using a full ab initio (AI) approach, based on time-dependent density functional simulations \citep{Gross} in the linear response (LR-TDDFT) \citep{POWELL197429}. \\
The so-derived dielectric functions are used as input for a Monte Carlo description of the inelastic scattering probability to calculate the energy loss
of electrons in their path within the solid. The comparison between our simulated and recorded REELS allows us to assess the impact that external tuneable parameters and semi-classical assumptions might have on the accuracy of simulated spectral lineshapes.


\section{Experimental details}
A polycrystalline diamond film was deposited on a silicon substrate in a microwave tubular reactor using a CH$_4$-H$_2$ gas mixture. After exposure to atmospheric pressure, the film was inserted into an Ultra High Vacuum (UHV) system equipped with a sample preparation chamber and an analysis chamber. Highly Oriented Pyrolytic Graphite (HOPG) was cleaved ex-situ before inserting into the UHV system. The two samples were cleaned by annealing at 600$^\circ$ C for 10 minutes in UHV. REEL spectra were acquired in a PHI 545 system operating at a base pressure of $\approx 2 \times 10^{-10}$ mbar. The instrument is equipped with a double-pass cylindrical mirror analyzer (CMA), a coaxial electron gun, a non-monochromatic MgK$\alpha$ ($h\nu$ = 1253.6 eV) X-ray source and a He discharge lamp. For a CMA, incoming electrons cross the surface at a fixed angle with respect to the sample normal, while outgoing electrons cross the surface at a variable angle dependent on the angle between the surface normal and the CMA axis (30$^\circ$), the entrance angle to the analyser (42$^\circ  \pm$ 6$^\circ$) and the azimuth angle in a plane normal to the CMA axis. Spectra are taken at a constant energy resolution of 0.6 eV, as measured on a Pd Fermi edge. The measured FWHM of the Zero Loss Peak (ZLP) is 0.9~ eV. The energy of incident electrons ranges from $T$ = 250 eV to $T$ = 2000 eV. Once acquired, REEL spectra are corrected for the energy dependence (E$^{-0.9}$) of the analyser transmission function.

\section{Computational details}

The physical quantity relating the microscopic and macroscopic description of electron beam interaction with matter is given by the dielectric response function. It is important to distinguish between microscopic and macroscopic quantities, where the latter are defined as averages over the unit cell of the former, because the total electric field caused by an external perturbing field can exhibit rapid oscillations at the atomic level, while at a larger scale the response function is homogeneus. 
For periodic crystals (as we model diamond and graphite films periodic in the in-plane direction), one can exploit the translational symmetry and the microscopic dielectric function is conveniently written in reciprocal space, i.e. $\epsilon_{\mathbf{G},\mathbf{G'}}(\mathbf{q},\omega) = \epsilon(\mathbf{q}+\mathbf{G}, \mathbf{q}+\mathbf{G'},\omega)$, where $\mathbf{G}$ and $\mathbf{G'}$ are reciprocal lattice vectors, $\mathbf{q}$ is the transferred momentum contained in the first Brillouin zone (IBZ) and $\omega$ is the transferred energy. With this notation $\epsilon_{\mathbf{G},\mathbf{G'}}(\mathbf{q},\omega)$ is also often called dielectric matrix. It can be shown \citep{ullrich} that the relation between the experimentally measurable macroscopic dielectric function and the microscopic one is \citep{PhysRev.126.413,PhysRev.129.62}
\begin{equation}\label{eM}
\epsilon_M( \mathbf{q}, \omega) =  \left [ \epsilon^{-1}_{\mathbf{G}=0,\mathbf{G'}=0}(\mathbf{q},\omega)  \right ]^{-1}
\end{equation}
In particular, electron transport observables such as the energy loss per unit path or the inelastic scattering cross section are proportional to the imaginary part of minus the inverse of the dielectric function \citep{Ritchie_PhysRev_1957}, which is called energy loss function (ELF):
\begin{equation}\label{REELS}
\text{ELF} = \text{Im} \left [- \frac{1}{\epsilon(\mathbf{q},\omega)} \right ]
\end{equation}
where we have removed the subscript M from $\epsilon$ to simplify the notation. 
In general the momentum transferred by electrons upon collisions is neither negligible nor constant in different energy ranges. Thus one needs to evaluate the dielectric function also out of the optical limit before calculating the expression in Eq. (\ref{REELS}). 
In this regard, the computation of the dielectric response for finite momentum transfer is a major issue in the treatment of inelastic interactions, and has a dramatic influence on energy 
loss spectra and secondary electron lineshapes. In the following sections we describe two different approaches at different level of accuracy to overcome this issue, 
notably the DL and the full AI models. \\
 We stress that the assumption behind the validity of Eq. (\ref{REELS}) is the first order Born approximation, which is reliable only for sufficiently fast charged particles that can thus be considered point-like and are weakly perturbed and deflected in each collision by the potential scattering.
 In general, the latter hypothesis means that the incident particle kinetic energy $T$ should exceed the one of target electrons, so that in our case $T\text{(eV)} \gg 13.6 Z^2\simeq 490$ eV.
 However, calculations using Eq. (\ref{REELS}) can be considered reliable well below this \citep{book_interaction}. Indeed, the latter value is referred to core-level electrons, while in our case only target valence electrons, having much lower kinetic energy, enter the model. 
We will see by comparison with our experimental data that quantitative results from Monte Carlo simulations based on Eq. (\ref{REELS}) can be obtained even for low-energy secondary electron yields. 

\subsection{Drude--Lorentz model}

The DL model approximates the material dielectric response to an applied uniform external field of frequency $\omega=\frac{E}{\hbar}$, with $E$ the perturbation energy and $\hbar$ the reduced Planck constant, by considering the target screening electrons as harmonic oscillators of frequency equal to the plasmon frequency $\omega_{n}=\frac{E_{n}}{\hbar}$, with $E_n$ the plasmon energy. Charge oscillation is damped via a damping constant $\Gamma_n$, that takes into account friction-like forces affecting the oscillatory harmonic motion.
Furthermore, within this approach the ELF is extrapolated outside the optical domain, for not-vanishing momentum, by using a quadratic dispersion law that resembles the Random Phase Approximation (RPA) \citep{book_interaction}. In the RPA, valence electrons in the solid are approximated by a non-interacting homogeneous gas, and the plasmon energy is expanded to the second order in $q$ as:

\begin{equation}\label{displaw}
E_n (q \neq 0) = E_n(q  = 0) + \frac{\hbar^2 q^2}{2m}
\end{equation}

The ELF can thus be expressed as a sum over all oscillators of $q$-dependent generalized DL functions with a full-width-half-maximum $\Gamma_n$ as follows \citep{Dapor_NIMB_2015,Yubero_PRB_1992,book_interaction}: 
\begin{equation}\label{DrudeELF}
\mathrm{Im} \left[ -\frac{1}{\epsilon(q, E)} \right] = \sum_{n} \frac{A_n\Gamma_n E}{{(E_n^2(q) - E^2)^2 - (\Gamma_n E)^2}}
\end{equation}
where $A_n$ is the oscillator strength of the  $n^{th}$-oscillator and can be obtained by fitting procedures of optical data. 
Most importantly, it can be shown that the $f$-sum rule is exactly satisfied by the Drude dielectric function.

\subsection{Ab initio simulations}

Ab initio simulations of diamond and graphite dielectric functions have been carried out using a TDDFT approach in linear-response (LR) approximation \citep{sottile}. 
TDDFT has shown its great potential to replace many-body perturbation methods for accessing excitation energies and spectra of solids in their interaction with electromagnetic fields. 
The key quantity to perform LR-TDDFT simulation is the polarization function $\chi(\mathbf{r},t,\mathbf{r'},t')$ describing 
the change of the density $\delta n$ at $(\mathbf{r},t)$ due to a small deviation in the external perturbation $\delta v_\mathrm{ext}$ at $(\mathbf{r'},t')$: 
\begin{equation}
    \delta n (\mathbf{r},t) = \int dt' \int d^3 r' \chi (\mathbf{r},t,\mathbf{r'},t') \delta v_\mathrm{ext} (\mathbf{r'},t')
\end{equation}
The TDDFT many-body response function $\chi (\mathbf{r},t, \mathbf{r'},t')$ is related to the independent-particle polarizability $\chi_{KS} (\mathbf{r},t,\mathbf{x},\tau)$
by a Dyson-type equation as follows:
\begin{equation}\small
\begin{split}
    \chi (\mathbf{r},t, & \mathbf{r'},t') =  \chi_\mathrm{KS} (\mathbf{r},t,\mathbf{r'},t') + \Big  \{ \int d \tau \int d^3 x \int d \tau' \int d^3 x' \\
&    \chi_\mathrm{KS} (\mathbf{r},t,\mathbf{x},\tau) \Big [  \frac{\delta(\tau - \tau')}{|\mathbf{x}-\mathbf{x'}|} + f_\mathrm{xc} (\mathbf{x},\tau,\mathbf{x'},\tau')  \Big ] \chi (\mathbf{x'},\tau',\mathbf{r'},t') \Big \}
\end{split}
\end{equation}
where $f_\mathrm{xc}(\mathbf{r},t,\mathbf{r'},t')=\frac{\delta v_\mathrm{xc}(\mathbf{r},t)}{\delta n (\mathbf{r'},t')} \Big |_{n_\mathrm{gs}(\mathbf{r},t)}$ is the energy-dependent exchange-correlation kernel.
The independent-particle response function $\chi_{KS}$ is calculated usually via a mean-field approach, such as Kohn-Sham DFT.  
As in static DFT, the time-dependent exchange-correlation potential is unknown and calculations usually rely on the so-called adiabatic local density approximation (ALDA) 
in which the energy-dependence of the functional is neglected \citep{Gross}. However, in systems where excitonic effects are expected to have a strong influence on spectral features 
due to an ineffective electronic screening, e.g. in insulators as diamond, the use of bootstrap kernel \citep{fxc210} that includes effects beyond the RPA is necessary.\\
Assuming translational invariance, the ELF can be computed inserting Eq. (\ref{eM}) into Eq. (\ref{REELS}). The inversion procedure can be cumbersome for large basis sets and large k-point grids. 
Thus, usually one calculates the so-called dielectric matrix without local field effects (LFE), in which the dielectric matrix off-diagonal elements ($\epsilon_{\mathbf{G},\mathbf{G'}}(\mathbf{q},\omega)$, $\mathbf{G},\mathbf{G'}\neq 0$) 
for different values of $\mathbf{q}$ are neglected \citep{PhysRev.129.62}. 
In the latter case, the macroscopic dielectric matrix is obtained by simply inverting the head of the microscopic dielectric matrix. 
However, in these off-diagonal terms the fluctuations on atomic scale of the polarization are encoded. Thus, for highly inhomogeneous electron density systems or highly locally polarizable atoms, such as in the case of diamond and graphite, LFE can play a significant role in the description of the dielectric properties \citep{physrevBvol61num15}, particularly at small wavelength, to the point of invalidating even qualitative results. Thus, in our analysis we include LFE in the dielectric properties of these carbon-based materials, where strong microscopic local fields can be created. 
Including LFE, the dielectric function in reciprocal space is \citep{kresse}:
\begin{equation}\label{diel}
    \epsilon^{-1}_{\mathbf{G},\mathbf{G'}}(\mathbf{q},\omega) = \delta_{\mathbf{G},\mathbf{G'}}+ \nu^{s}_{\mathbf{G},\mathbf{G'}}(\mathbf{q}) \chi_{\mathbf{G},\mathbf{G'}}(\mathbf{q},\omega)
\end{equation}
where $\nu^{s}_{\mathbf{G},\mathbf{G'}}(\mathbf{q})=\frac{4 \pi e^2}{|\mathbf{q}+\mathbf{G}||\mathbf{q}+\mathbf{G'}|}$ is the Fourier transform of the Coulomb potential, and
$\chi_{\mathbf{G},\mathbf{G'}}(\mathbf{q},\omega)$ is the microscopic polarizability.\\
The computation of the dielectric and energy-loss functions within the TDDFT framework has been carried out by using the ELK code, which is an all-electron full-potential linearised augmented-plane-wave (FP-LAPW) program \citep{elk}.
More specifically, the electron-electron interaction in diamond crystals was described via the generalized gradient approximation (GGA) \citep{gga20} using a "bootstrap" exchange-correlation kernel $f_\mathrm{xc}$ \citep{fxc210}, with the cut-off for the augmented plane waves set to $665$ eV.  
Brillouin zone sampling was performed using a $20 \times 20 \times 20$ $k$-point grid along with an electron occupancy Fermi smearing of $0.2$ eV, which ensures convergence of the dielectric observables 
for the unitary cell below chemical accuracy.\\
In the case of graphite, a local spin-density approximation (LSDA) \citep{PhysRevB.45.13244} to the exchange-correlation functional has been used together with an ALDA for the exchange-correlation kernel $f_\mathrm{xc}$. The Brillouin zone was sampled using a $16 \times 16 \times 16$ $k$-point grid, while the other DFT parameters were kept the same as for diamond. 

\subsection{Theory of Monte Carlo simulations}\label{MC}
The transport of electrons within a material can be simulated by a classical MC approach, assuming that the non-relativistic electron beam wavelength is small with respect to interatomic separation \citep{book_interaction} and that the scattering cross sections for the different processes occurring within materials are known. \\
At this level the target is assumed to be semi-infinite, homogeneus and amorphous, the latter conditions supporting the assumption of incoherent scattering between different events. 
In our transport model, we consider a mono-energetic electron beam with $N$ electrons impacting on the target with  kinetic energy $T$ and angle of incidence $\theta$ with respect to surface normal. \\
Electrons can undergo elastic and inelastic scattering. The scattering is usually elastic when electrons scatter nuclei with far heavier mass, and only a trajectory change by an angle $\theta$ is recorded. In this case the elastic cross section $\sigma_{el}$ is calculated by using the Mott theory, which is based on the solution of the Dirac equation in a central field \citep{Dapor_book_bianco}. At variance, inelastic scattering processes resulting into both an energy loss $W$ and a directional change $\theta$ are mainly due to electron--electron interactions.
In this occurrence, the inelastic cross section $\sigma_\mathrm{inel}$ is assessed via the dielectric model \citep{Ritchie_PhysRev_1957}. 
Within this approach the differential inelastic cross section is calculated as follows:
\begin{equation}
\frac{d\sigma_\mathrm{inel}}{dW} = \frac{1}{\rho \pi a_0 T} \int_{q_-}^{q_+} \frac{dq}{q} \mathrm{Im}\left[- \frac{1}{\epsilon (q,W)}\right]
\label{sinel_diff}
\end{equation}
where $a_0$ is the Bohr radius, $\rho$ the atomic density of the target material, $q$ is the transferred momentum and the integration limits are $q_- = \sqrt{2m}(\sqrt{T} - \sqrt{T-W})$ and $q_+ = \sqrt{2m}(\sqrt{T} + \sqrt{T-W})$. \\
The electron mean free path $\lambda$ is given by \citep{Dapor_book_bianco}:
\begin{equation}
\lambda = \frac{1}{\rho (\sigma_\mathrm{el} + \sigma_\mathrm{inel})}
\end{equation}
Furthermore, we assume that the path travelled by a test charge between two subsequent collisions is Poisson--distributed, so that the cumulative probability that the electron moves a distance $\Delta s$ before colliding is given by:
\begin{equation}
\Delta s = - \lambda \cdot \mathrm{ln}(r_1)
\end{equation}
The random numbers $r_1$, as well as all random numbers employed in our MC simulations, are sampled in the range [0,1] with an uniform distribution.
A second random number $r_2$ is compared with the elastic ($p_\mathrm{el} = \frac{\lambda_\mathrm{el}}{\lambda\mathrm{el} + \lambda\mathrm{inel}}$) and inelastic ($p_\mathrm{inel} = 1-p_\mathrm{el}$) scattering probability to 
determine weather the scattering is elastic ($r_2 < p_\mathrm{el}$) or inelastic.
The outcome of an elastic interaction is given by the trajectory deflection of an angle $\theta'$ with respect to the direction before the collision, which can be computed by equalizing the following cumulative elastic
probability with a third random number $r_3$: 
\begin{equation}
P_\mathrm{el}(\theta',T) = \frac{1}{\sigma_\mathrm{el}} 2 \pi \int_0^{\theta'} \frac{d\sigma_\mathrm{el}}{d\theta} d\theta= r_3 
\end{equation}
On the other hand, inelastic processes are dealt with by computing the inelastic scattering probability as:
\begin{equation}\label{pinel}
P_\mathrm{inel}(W, T) =  \frac{1}{\sigma_\mathrm{inel}}\int_0^{W} \frac{d\sigma_\mathrm{inel}}{dW'} dW'
\end{equation}
As customary in electronic transport MC calculations the maximum energy loss corresponds to half of the kinetic energy of the incident electron. To determine the energy loss $W$, 
we generate a database of $P_\mathrm{inel}$ values for different $W$ and $T$ and we equalize the integral in Eq. (\ref{pinel}) to a random number $r_4$. \\
Eventually, scattered electrons can be ejected from the target. In this work, we will calculate the so-called secondary emission yield, which is defined as
the ratio between the number of secondary electrons emitted from the surface and the number of electrons in the beam prior the scattering.
The assessment of the secondary electron spectral features is particularly important in imaging techniques \citep{Dapor_book_blu,Dapor_APSUSC_2015}. 
In their way out of the solid the electrons lose further energy to overcome the potential barrier $E_{\text A}$ (electron affinity) at the interface of the material. This process can be modelled as scattering by a potential barrier. Thus, the transmission coefficient can be computed as follows: 
\begin{equation}
t = \frac{4 \sqrt{(1 - E_{\text A}/(T \cdot \mathrm{cos}^2\theta'_z))}}{(1 + \sqrt{(1 - E_{\text A}/(T \cdot \mathrm{cos}^2\theta'_z)) })^2}
\end{equation}
where $t$ represents the probability that the electron leaves the sample's surface and $\theta_z'$ is the incident angle  with respect to the surface normal. Finally, by comparing this transmission coefficient with a random number $r_5$, electrons are (or are not) emitted into the continuum with a kinetic energy lowered by the work function $E_\mathrm{A}$ whenever $t>r_5$ ($t<r_5$). By definition, emitted electrons emerging with kinetic energies below 50 eV are called secondary electrons and their spectral features.
The Monte Carlo routines used for performing these simulations are embedded in the in-house suite SURPRISES \citep{1749-4699-2-1-015002,taioli2009mixed}.


\section{Results and discussions}

\subsection{Frequency-dependent dielectric function of diamond and graphite}

\begin{figure}[ht!]
\centering
\includegraphics[width=0.48\linewidth]{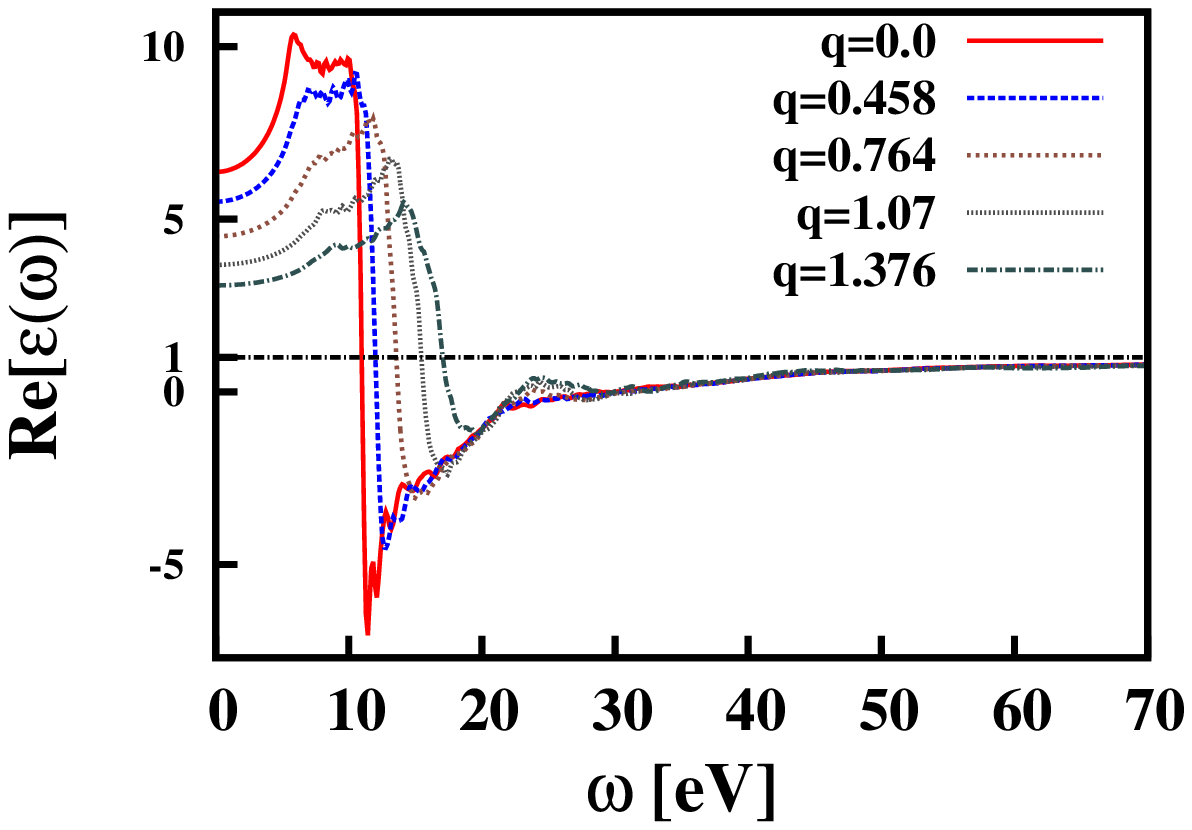}
\includegraphics[width=0.48\linewidth]{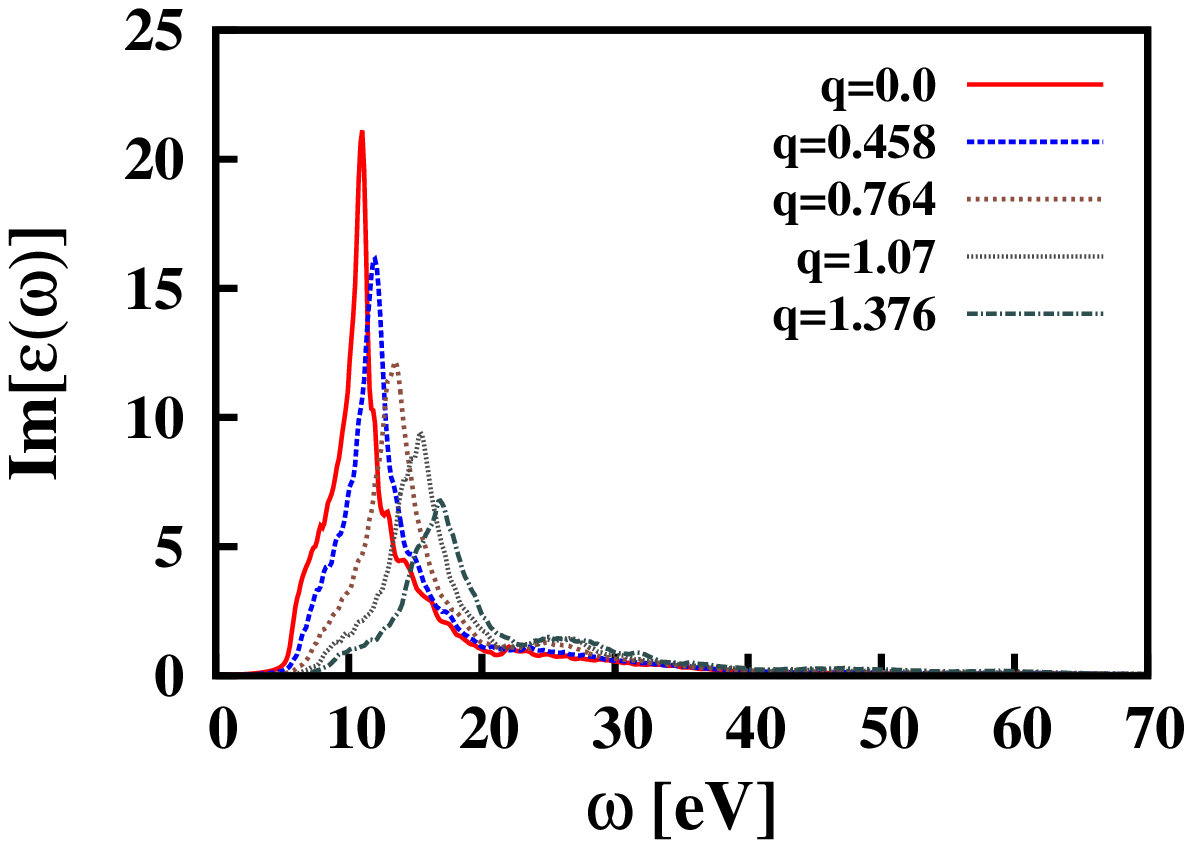}
\caption{Real (left panel) and imaginary (right panel) parts of the dielectric function, obtained from ab initio calculations, of diamond vs. energy transfer (eV) for different momentum transfer q ($\AA^{-1}$) along the $\Gamma$L direction.}
\label{fig:epsilon_diamond}
\end{figure}

\begin{figure}[ht!]
\centering
\includegraphics[width=0.48\linewidth]{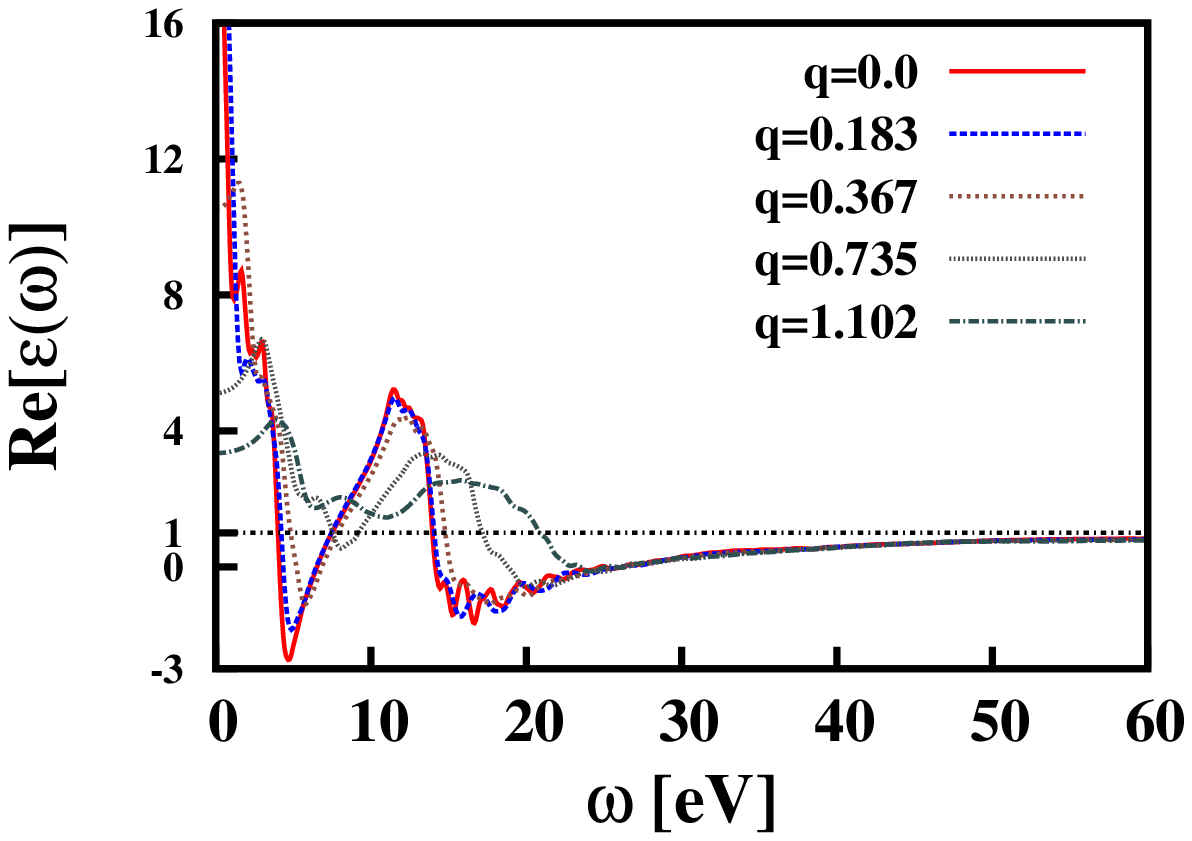}
\includegraphics[width=0.48\linewidth]{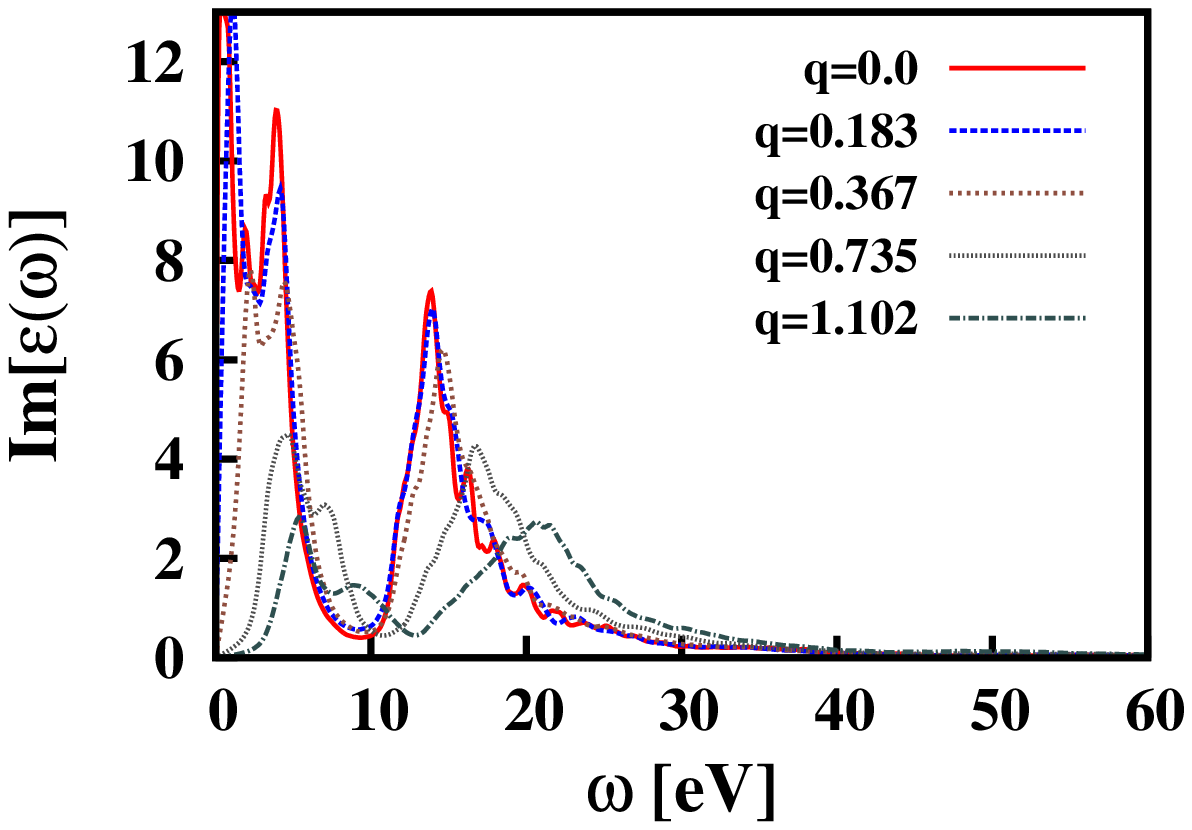}
\caption{Real (left panel) and imaginary (right panel) parts of the dielectric function,obtained from ab initio calculations, of graphite vs. energy transfer (eV) for different momentum transfer q ($\AA^{-1}$) along the $\Gamma$M direction.}
\label{fig:epsilon_graphite}
\end{figure}

The central quantity of our analysis is represented by the dielectric function, which links the microscopic features of diamond and graphite, accessible from ab initio or model simulations,
to REELS experimental data. The real and imaginary part of the frequency-dependent dielectric function of diamond and graphite from our ab initio simulations are reported respectively in Figs. \ref{fig:epsilon_diamond} and \ref{fig:epsilon_graphite} for several transferred momenta in the IBZ along the $\Gamma$L direction for diamond and along the $\Gamma$M direction for graphite.
We notice that both real and imaginary parts of $\epsilon$ are strongly dependent on the transferred momentum, showing a high degree of anisotropy in the dielectric response for both solids. Furthermore, we stress that the dielectric function real and imaginary part expected asymptotic behaviour (that is the real part goes to one, while the imaginary part goes to zero for increasing energy transfer at any $\mathbf{q}$) is rigorously satisfied for both solids.

\subsection{Frequency-dependent energy loss function of diamond and graphite}

The ELF can be accurately computed via Eq. (\ref{REELS}) within a full ab initio model once the dielectric function appearing in Eq. (\ref{diel}) is known in a fine grid of $q$ points.
A computationally less expensive approach could be used, in which the dielectric function in the optical limit is computed from ab initio simulations or taken from photo-emission experimental 
data \citep{exp_diam,exp_grafite} while the extension to finite momentum transfer is obtained by applying a RPA-type dispersion law. 
In this regard, to test the accuracy of different models we calculated the ELF by three approaches.\\

\begin{figure}[ht!]
\centering
\includegraphics[width=0.98\linewidth]{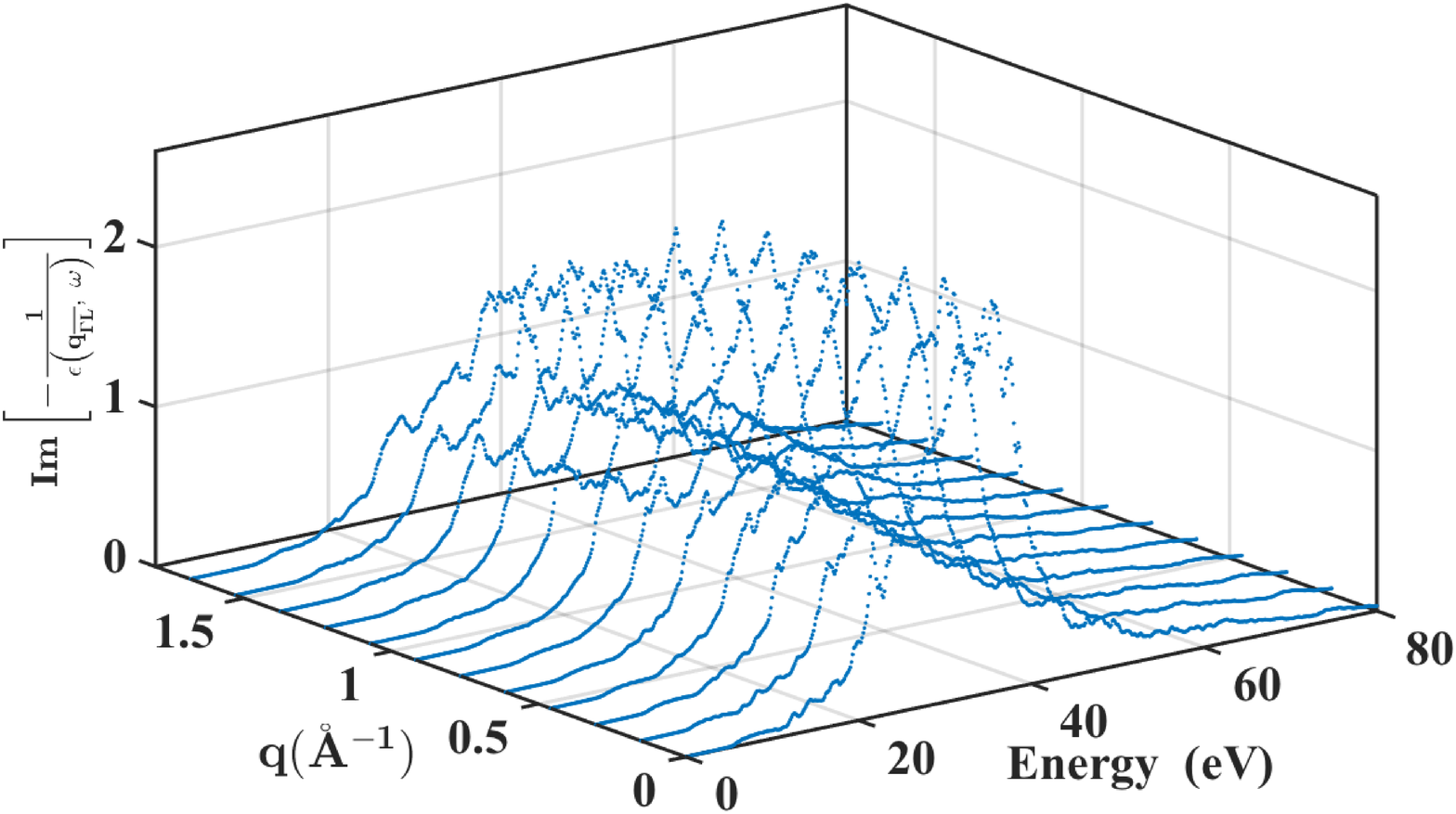}
\caption{ELF of diamond, obtained from ab initio simulations vs. energy (eV) and momentum ($\AA^{-1}$) transfer along the IBZ $\Gamma$L direction.}
\label{fig:diam_back}
\end{figure}

\begin{figure}[ht!]
\centering
\includegraphics[width=0.98\linewidth]{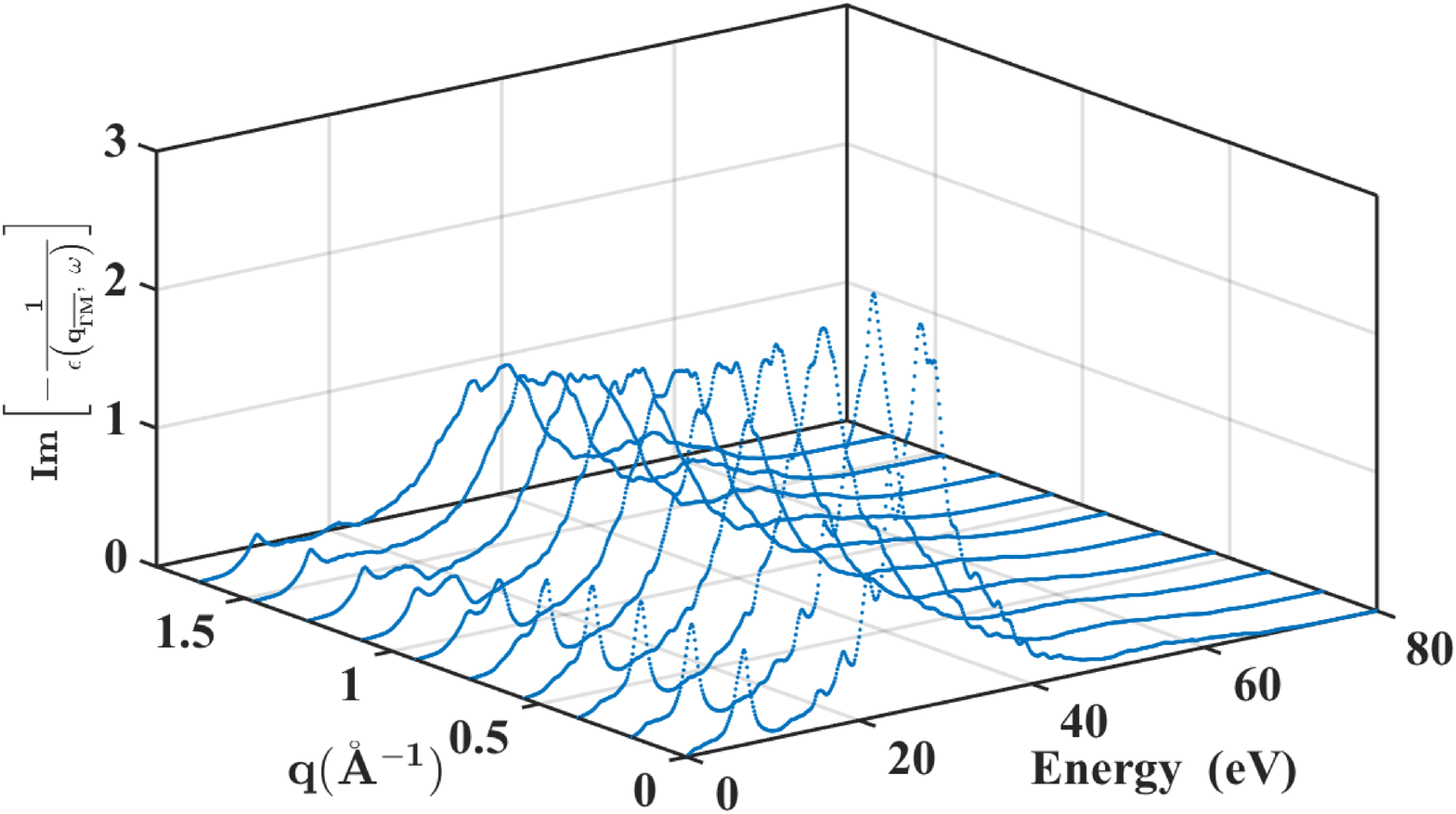}
\caption{ELF of graphite, obtained from ab initio simulations vs. energy (eV) and momentum ($\AA^{-1}$) transfer along the IBZ $\Gamma$M direction.}
\label{fig:graph_back}
\end{figure}

{\it Full ab initio} (AI): Within this approach, the ELF was computed with a full AI approach both for $\mathbf{q} \to 0$ and outside the optical limit. 
The ELF, plotted as a function of energy and momentum transfer along the $\Gamma$L ($\Gamma$M) direction for diamond (graphite), are shown in Figs. \ref{fig:diam_back} and \ref{fig:graph_back}, respectively. \\

\begin{figure}[ht!]
\centering
\includegraphics[width=0.48\linewidth]{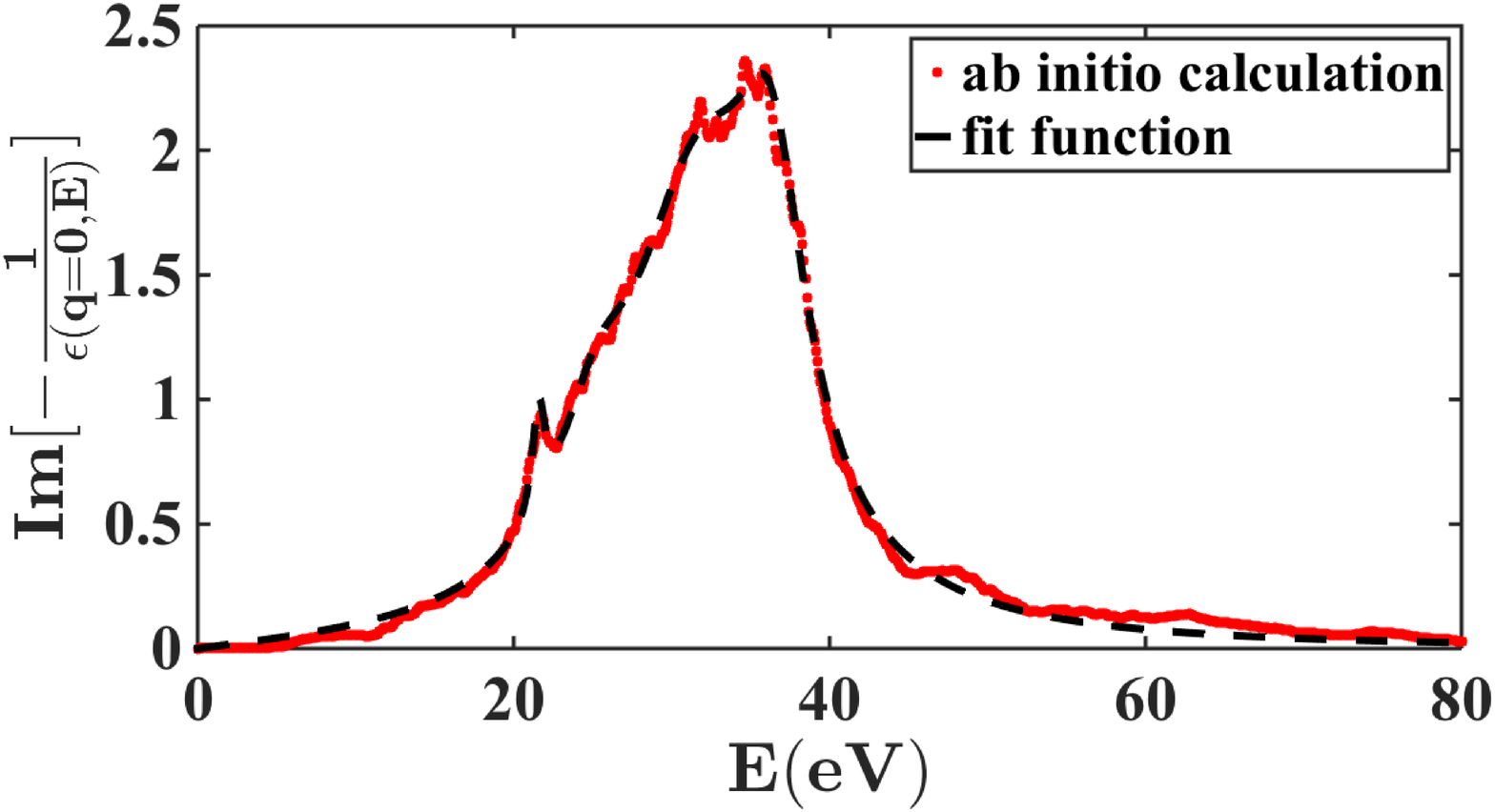}
\includegraphics[width=0.48\linewidth]{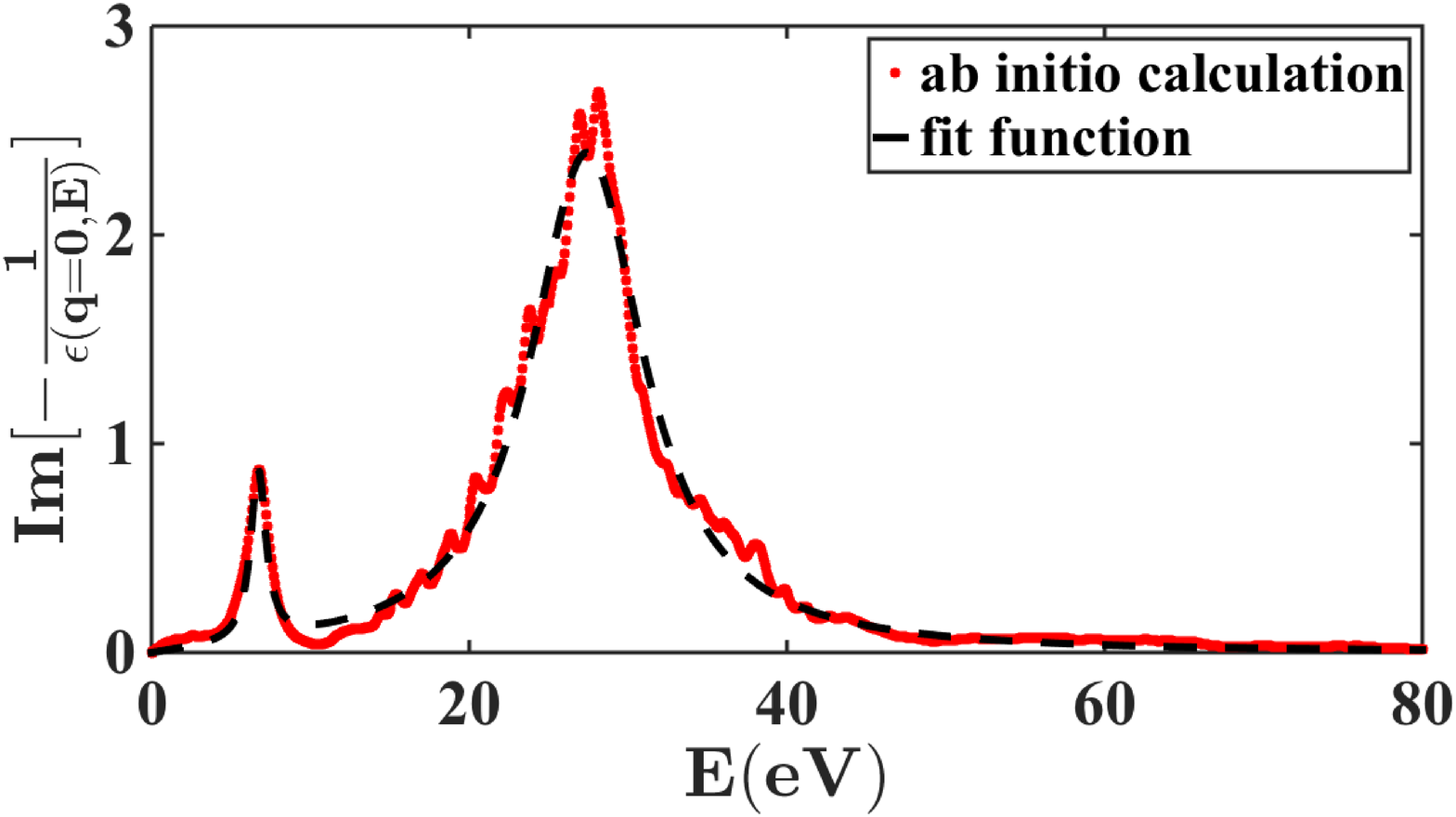}
\caption{ELF in the optical limit ($\omega \rightarrow 0$) obtained from AI simulations (continuous red curve) along with the data fit (dashed black line) for diamond (left panel) and graphite (right panel).}
\label{fig:elfFIT}
\end{figure}

{\it Drude--Lorentz from AI optical data} (DL--AI): Within this approach, the dielectric response in the optical domain was still obtained from AI simulations of the dielectric function, and the ELF computed using Eq. (\ref{displaw}) and then fitted via DL-functions as by Eq. (\ref{DrudeELF}). 
The extension to finite transfer momentum was finally achieved by using the dispersion law for plasmons reported in Eq. (\ref{displaw}). 
In the case of diamond four harmonic oscillators were sufficient to obtain an optimal fit of the AI optical data, while for graphite only two harmonic oscillators were used.
In Fig. (\ref{fig:elfFIT}) we compare the AI dielectric response (red curve) for diamond (left panel) and graphite (right panel) obtained in the optical limit with the DL fit (dashed black line).

The fitting parameters $E_n$, $\Gamma_n$ and $A_n$ have a clear physical meaning, representing respectively the plasmon peak energies, the damping constants characterizing the 
finite life-time of the quasi-particle excitation, and the oscillator strengths. The parameters obtained from this fitting procedure are reported in Tab. \ref{tab:fit} for diamond and graphite, respectively. \\
\begin{table}[ht!]
\centering
\begin{tabular}{c c c c}
     \hline
     n & $E_n$ (eV) & $\Gamma_n$ (eV) & $A_n$(eV$^2$)\\
     \hline
     Diamond\\
     \hline
     1 & 21.59 & 0.95 & 8.58 \\
     2 & 25.40 & 5.68 & 61.62 \\
     3 & 32.28 & 11.38 & 626.46\\
     4 & 36.39 & 5.25 & 224.76 \\
     \hline
     Graphite\\
     \hline
     1 & 6.75 & 1.17 & 6.38 \\
     2 & 27.76 & 8.68 & 573.09 \\
     \hline
\end{tabular}
\caption{Fitting parameters obtained by fitting the AI ELF in the optical limit with DL functions for diamond and graphite, respectively.}
\label{tab:fit}
\end{table}
We notice that our fit functions satisfy the $f$-sum rule \citep{Shiles_PRB_1980}, stating that the integral of the ELF multiplied by the energy loss sums up to the number of electrons per atom. For both diamond and graphite we obtained 3.7, a value close to 4 that is the number of electrons populating the $s$ and $p$ orbitals of the carbon atom. \\

\begin{figure}[ht!]
\centering
\includegraphics[width=0.48\linewidth]{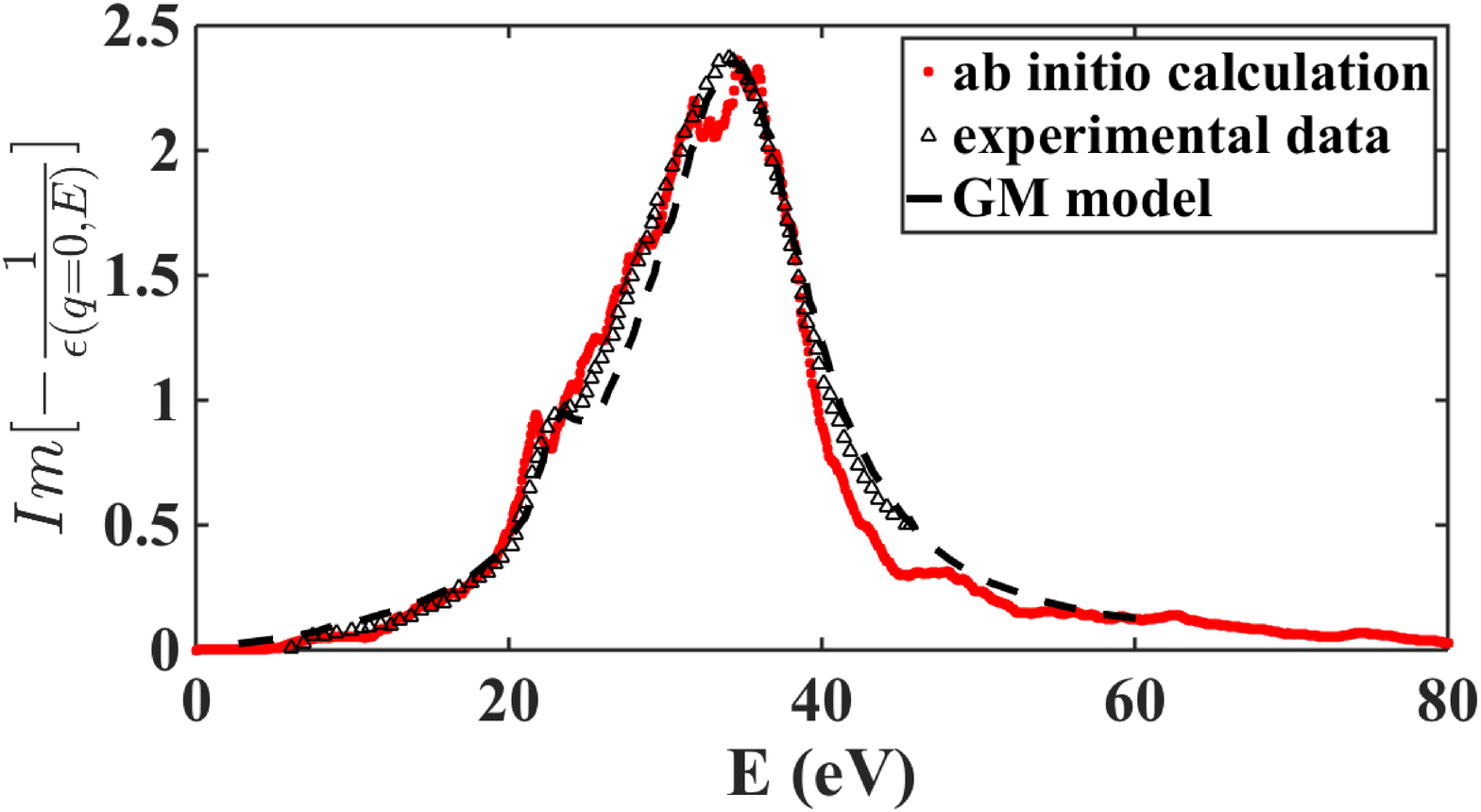}
\includegraphics[width=0.48\linewidth]{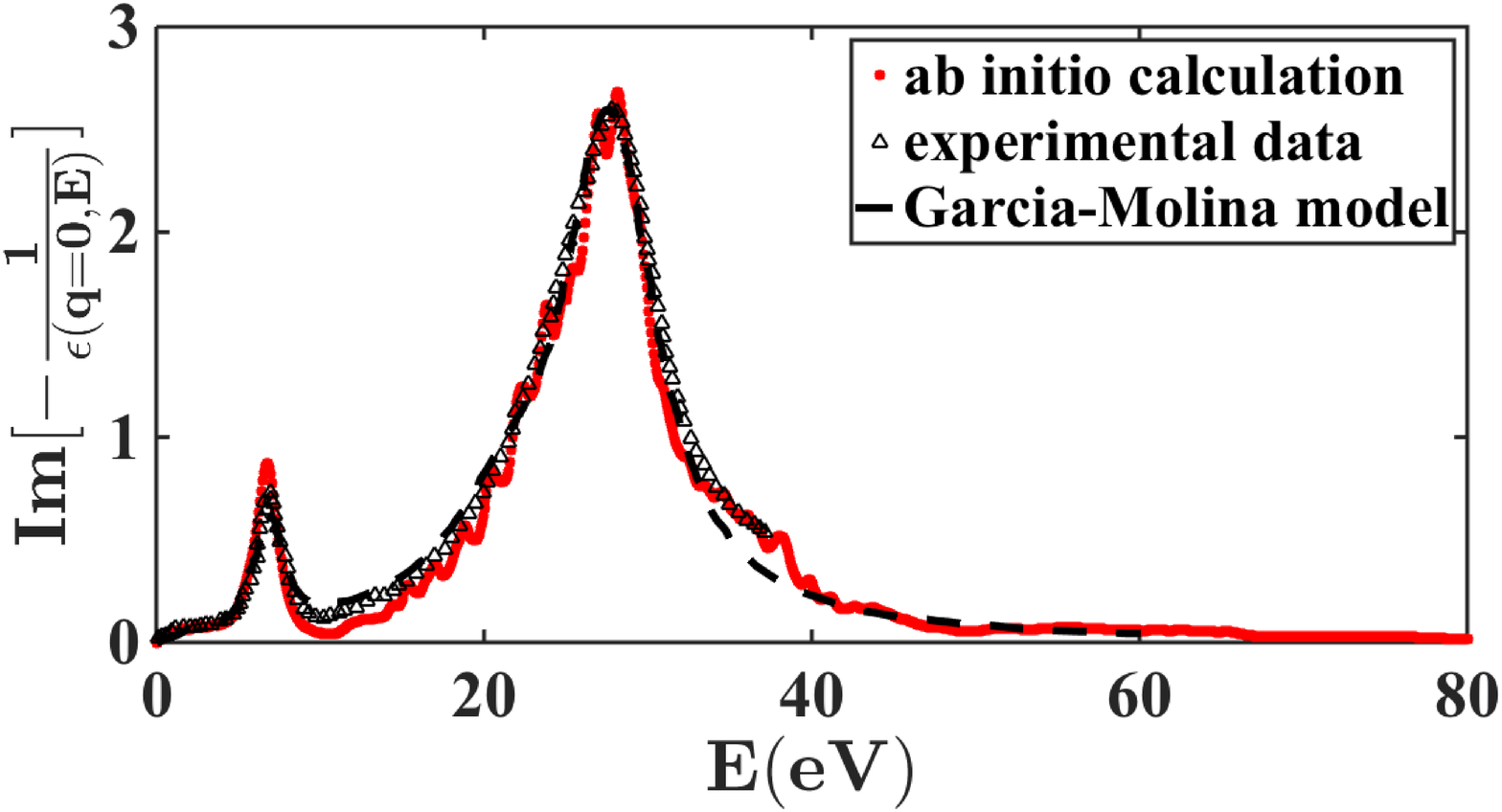}
\caption{Comparison between the ELF of diamond (left panel) and graphite (right panel) in the optical limit obtained from AI simulations (continuous red curve), experimental data from Refs. \citep{exp_diam} and \citep{exp_grafite} (black triangles) and fit obtained with the model of Garcia-Molina et al. \citep{Molina_NIMB_2006} (dashed black line).}
\label{fig:cfrELF}
\end{figure}

{\it Drude--Lorentz from experimental optical data} (DL--E): The ELF in the optical domain can be directly measured in photo-emission experiments, 
fitted with DL functions as reported by Garcia-Molina {\it et al.} \citep{Molina_NIMB_2006} and extended out of the optical limit by applying the dispersion law  reported in Eq. (\ref{displaw}).\\
In Fig. \ref{fig:cfrELF} we compare the ELF of diamond (left panel) and graphite (right panel) obtained from our AI simulations with the experimental data from Refs. \citep{exp_diam} and \citep{exp_grafite} and the fit model by Garcia-Molina {\it et al.} \citep{Molina_NIMB_2006}. We notice that Garcia-Molina et al. used a different number of oscillators, plasmon peak energies and FWHM to accurately fit the experimental data with respect to our model derived from AI simulations. 

\subsection{Inelastic Mean Free Path and Stopping Power}

\begin{figure}[ht!]
\centering
\includegraphics[width=0.48\linewidth]{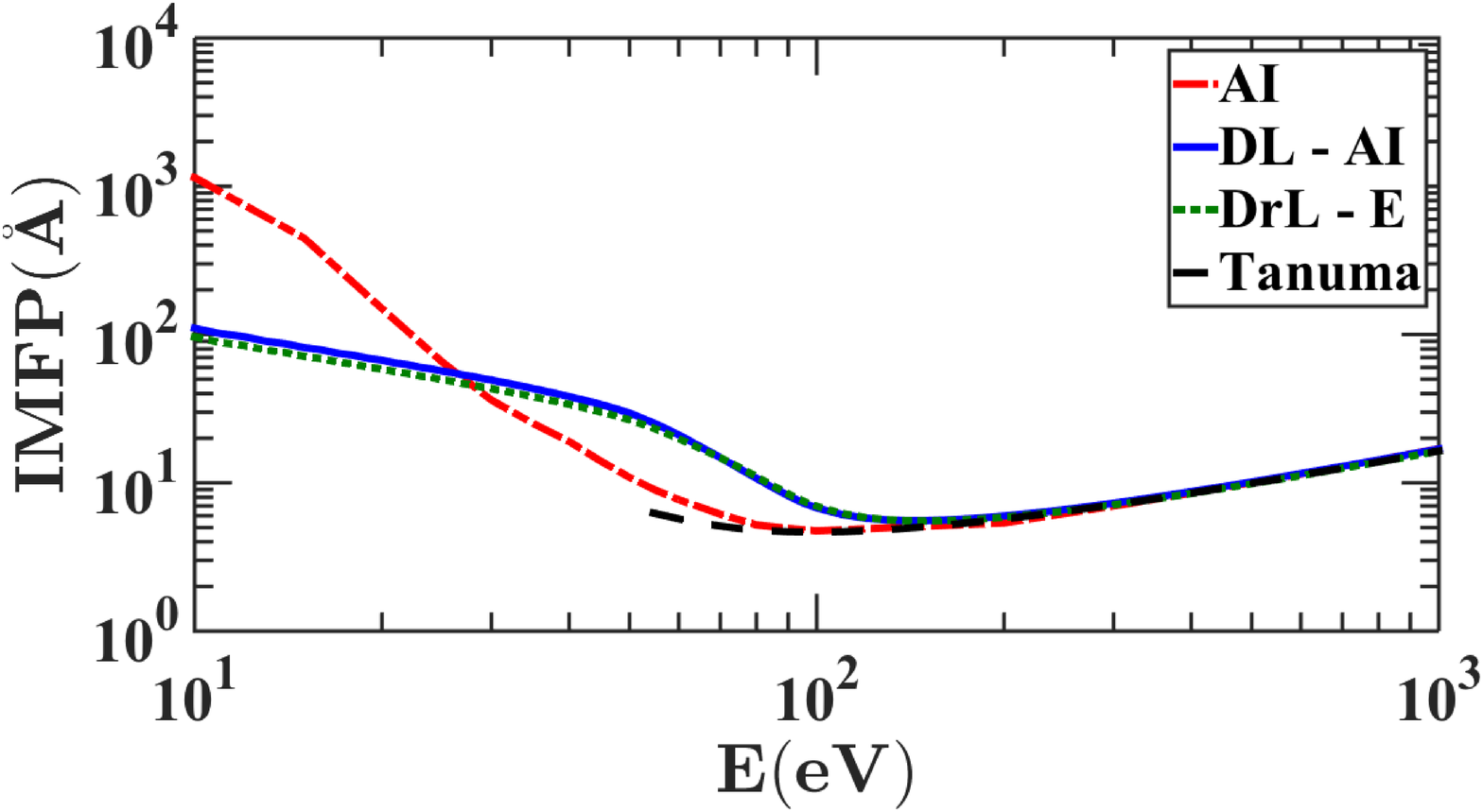}
\includegraphics[width=0.48\linewidth]{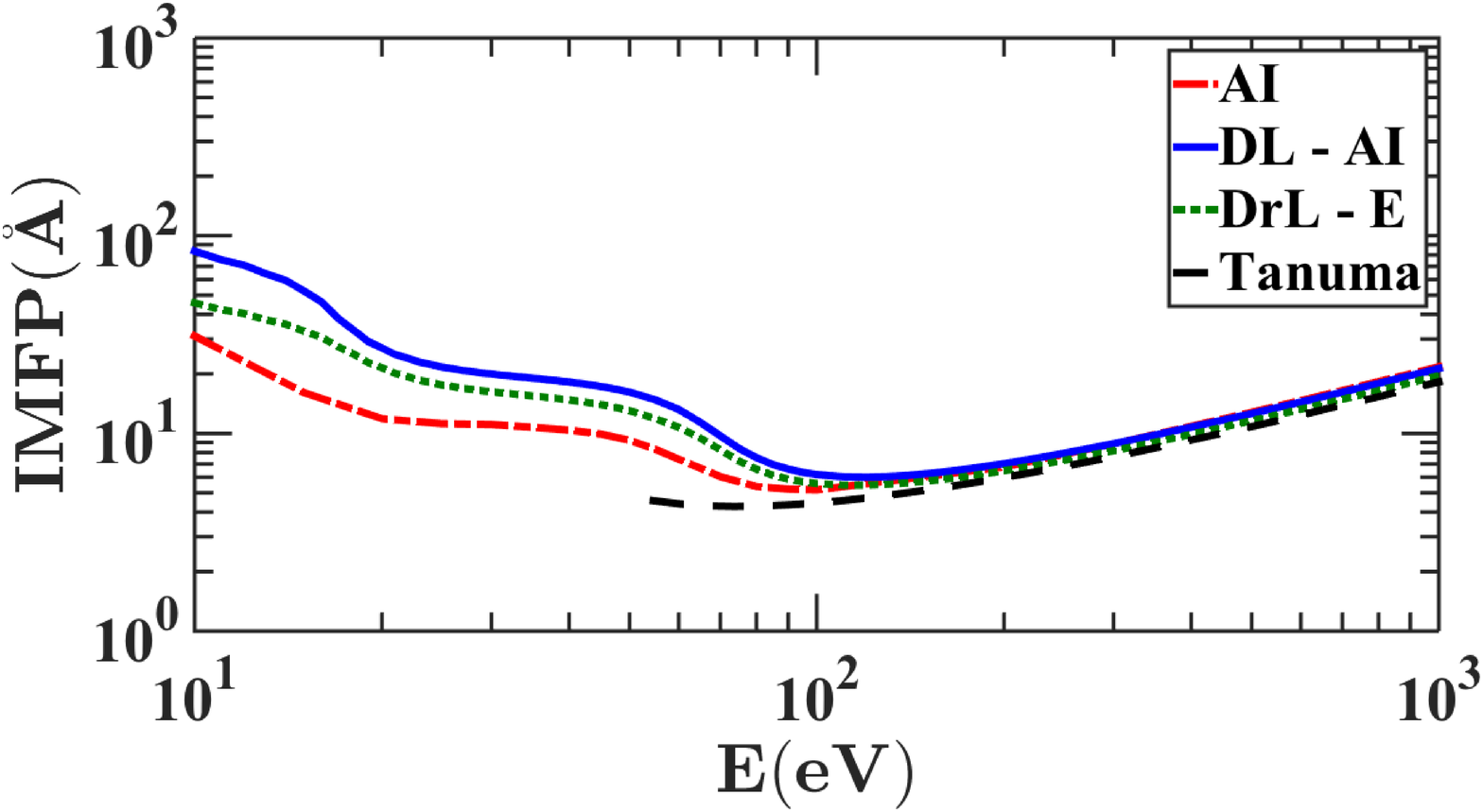}
\caption{IMFP of diamond (left) and graphite (right). Data obtained from AI simulations are reported in red (AI), DL-AI in blue and with the DLE in green. Black dashed lines correspond to the same quantities obtained by Tanuma {\it et al.} \citep{Tanuma_SIA_2009}.} 
\label{fig:IMFP_dg}
\end{figure}

\begin{figure}[ht!]
\centering
\includegraphics[width=0.48\linewidth]{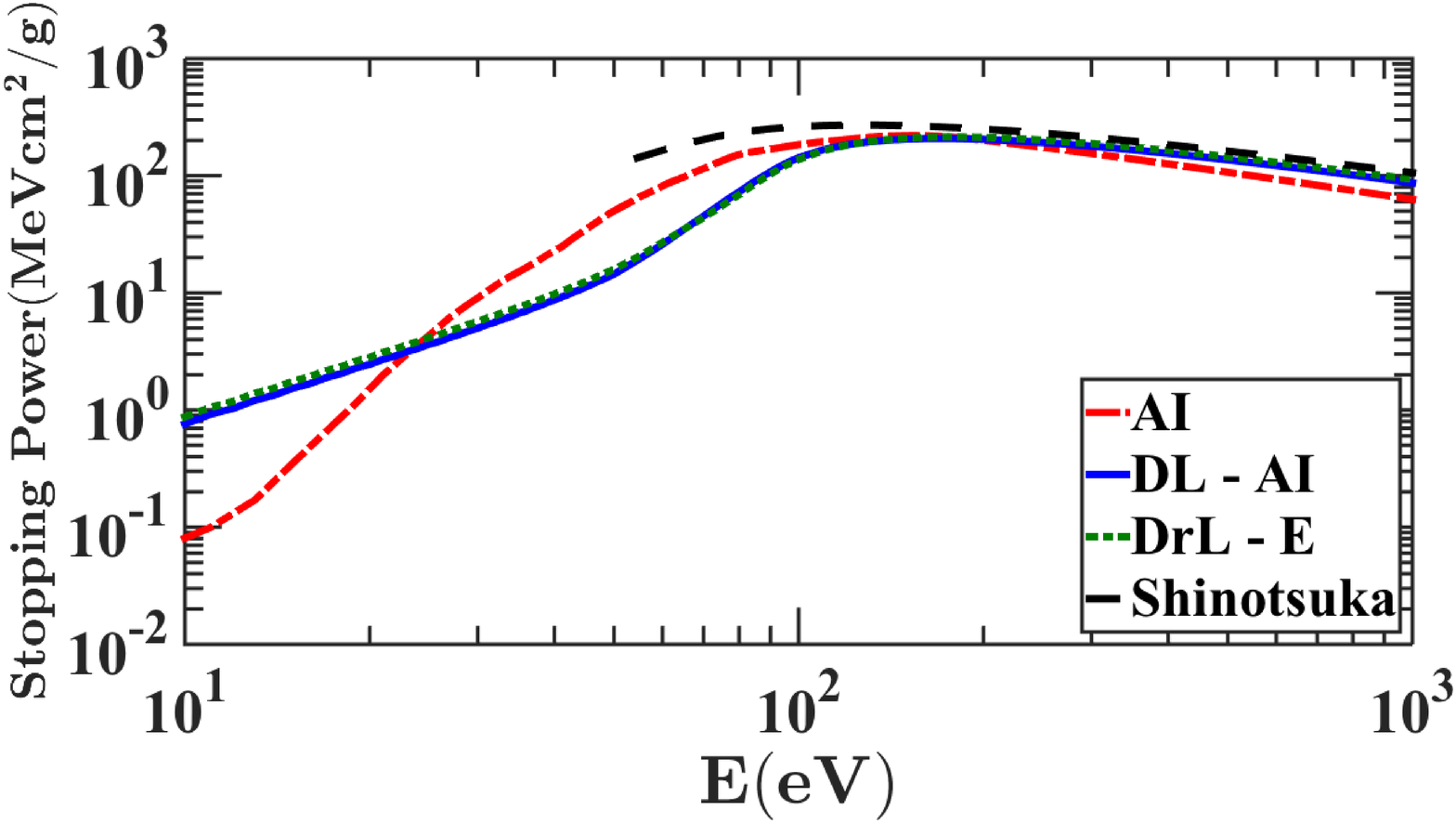}
\includegraphics[width=0.48\linewidth]{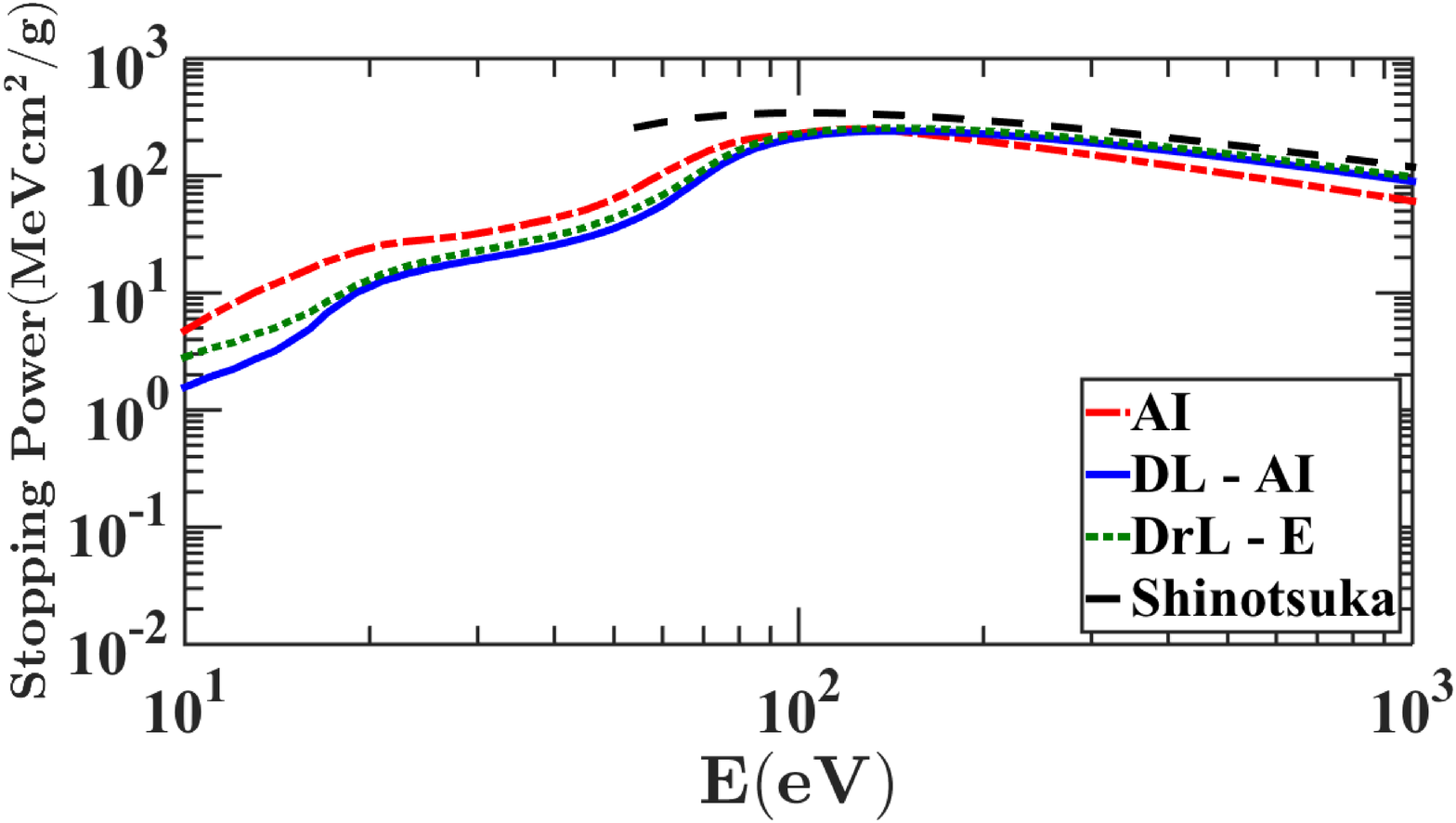}
\caption{Stopping power of diamond (left panel) and graphite (right panel). Data obtained with the AI approach are reported in red, DL-AI in blue, DL-E in green. Results from Shinotsuka et al. \citep{Shinotsuka} are sketched using a black dashed line.}
\label{fig:std}
\end{figure}

Using these different models of ELFs, we calculated a number of measurable physical quantities, such as the inelastic cross section (ICS), the inelastic mean free path (IMFP) and the stopping power (SP). These quantities will be used as input to our Monte Carlo calculation of electron energy 
loss spectra and yields of emitted secondary electrons.
We remember that the IMFP ($\lambda_\mathrm{inel}$), ICS ($\sigma_\mathrm{inel}$) and the SP ($S(T)$) are given by 
\begin{equation}
\lambda_\mathrm{inel} = \frac{1}{\rho \sigma_\mathrm{inel}},\quad \sigma_\mathrm{inel} = \int_{E_g}^{\frac{T+E_\mathrm{g}}{2}} \frac{d\sigma_\mathrm{inel} }{dW} dW,\quad S(T) = \int_{E_\mathrm{g}}^{\frac{T+E_\mathrm{g}}{2}} \frac{\lambda_\mathrm{inel}^{-1}}{dW} W dW
\end{equation}
where $\rho$ is the atomic density of the material. The upper and lower integration limits are, conventionally, fixed to the energy gap $E_\mathrm{g}$ and one half of the initial kinetic energy $T$ plus the energy band gap $E_\mathrm{g}$.  
In Fig. \ref{fig:IMFP_dg} we plot the IMFP of diamond (left) and graphite (right) 
calculated according to the three approaches mentioned above. Calculations according to the Tanuma--Powell--Penn (TTP) model \citep{Tanuma_SIA_2009} are also shown for reference.
The agreement between the different approaches and Tanuma's data is remarkably good, for both diamond and graphite, at energy transfers higher than 100 eV. However,
for energies below $100$ eV we find a significant discrepancy between the IMFP obtained by full AI calculations and the IMFP obtained via the DL dispersion law of Eq. (\ref{displaw}) (about one order of magnitude in the case of diamond). The large discrepancy will affect also our Monte Carlo simulations: an AI IMFP bigger than one order of magnitude at low energy with respect to DL models means that electrons have significantly lower probability to undergo inelastic scattering within the target material.  
One can explain this substantial difference between AI and DL models as due to the application of the quadratic dispersion law of Eq. (\ref{displaw}). In fact, this dispersion law is obtained using an homogeneous electron gas model within a RPA framework, which fails in the case of wide band-gap semiconductors or insulators, such as diamond. On the other hand this approximation works better for graphite, which shows an almost semi-metallic behaviour along the in-plane direction. \\
Finally, the SP obtained by using the three different approaches is reported in Fig. \ref{fig:std} for diamond (left panel) and graphite (right panel) and is compared with previous simulations by 
Shinotsuka et al. \citep{Shinotsuka}. 
Discrepancies among the different approaches at energies below 100 eV are evident also in this case. 

\section{Monte Carlo simulations}

\subsection{Diamond energy loss spectra and secondary electron yield}

\begin{figure}[ht!]
\centering
\includegraphics[width=0.98\linewidth]{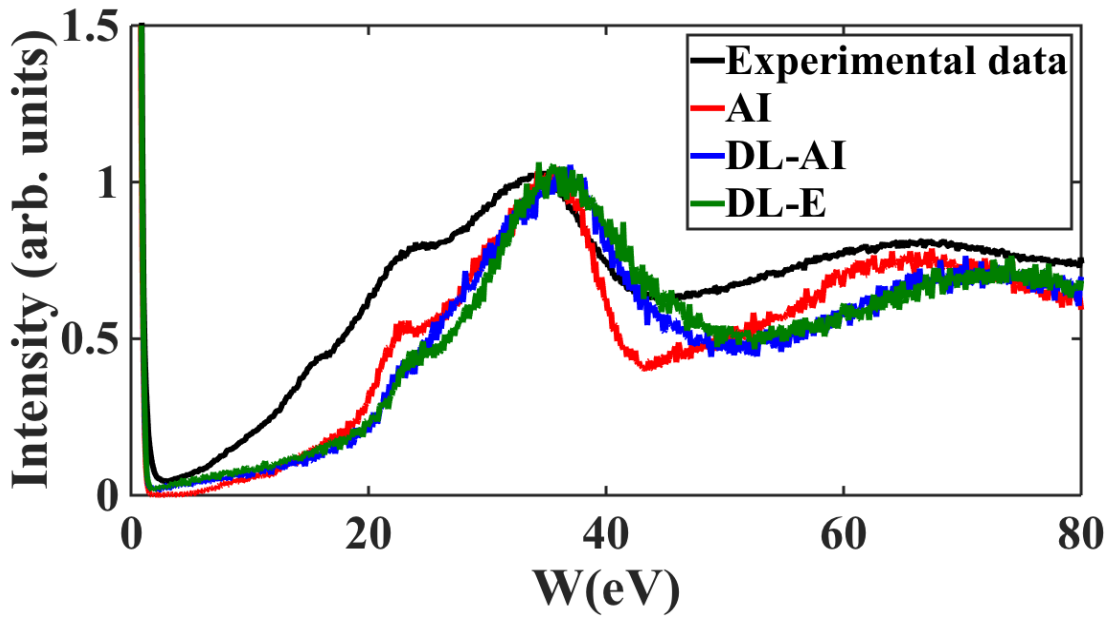}
\caption{REELS of diamond: experimental data are reported in black, while simulated results using the three different dielectric models are sketched in red (AI), blue (DL-AI) and green (DL-E). Electron beam kinetic energy is $1000$ eV. Data are normalized with respect to the $\sigma$ plasmon peak.}
\label{fig:pl_d}
\end{figure}

To compare our experimental REELS of diamond with the three different models of ELF presented above, we performed Monte Carlo (MC) simulations following the scheme reported in section \ref{MC}. In our MC simulations, diamond crystals are approximated by a homogeneous system with density $3.515$ g/cm$^3$ \citep{Molina_NIMB_2006}. 
Consequently, one assumes also that the ELF is almost similar in all directions and thus we can retain our simulated ELF along the $\Gamma$L direction for calculating the energy loss spectra. 
The band gap of diamond was set equal to $4.16$ eV, to be consistent with our DFT ab initio calculations.
The electron beam direction is orthogonal to the target surface and the initial kinetic energy ranges from 250 to 2000 eV. The number of impinging electrons is $10^9$. \\
In Fig. \ref{fig:pl_d} spectra of backscattered electrons simulated in terms of the three different models of the ELF are compared with our REELS experimental data. Simulated and experimental spectra present the $\sigma$ plasmon peak at $\sim 35$ eV, related to the four valence electrons of the equivalent covalently bonded carbon atoms. This finding is in agreement with the ELF function of Fig. \ref{fig:diam_back}, showing a maximum at about the same energy.

Furthermore, the two-plasmon excitation at higher energy ($\sim 70$ eV) in the experimental spectrum is also present in our MC simulations. We observe that while the MC simulations carried out using the 
dispersion law of Eq. (\ref{displaw}) show a blue shift with respect to experimental data, the use of a full AI approach results in a better agreement with experiments. 
We can conclude that at least in the case of insulators, due to the strongly dishomogeneous electron density and, thus, to the complexity of the dielectric response, the DL model is quantitatively less accurate than a full AI approach in the prediction of the experimental REELS. 
This behaviour worsens at higher transferred momenta, where particle-hole excitations, rather than collective plasma excitation, come into play. Single particle excitations generally cannot be well described by a simple RPA or by the DL model of the ELF, while TDDFT AI simulations are able to take into account also these spectral features.  

\begin{figure}[ht!]
\centering
\includegraphics[width=0.7\linewidth]{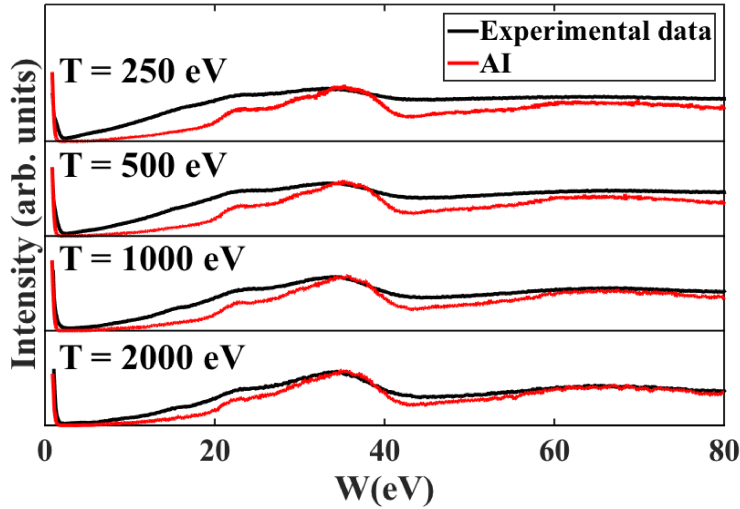}
\caption{REELS of diamond obtained by calculating the ELF with the full AI approach (red line) at different momentum transfer compared to experimental data (black line) for several primary electron beam kinetic energies $T$. Data are normalized with respect to the $\sigma$ plasmon peak.}
\label{fig:s_d_t}
\end{figure}

REEL experimental spectra and MC simulated spectra obtained using the full $AI$ method are compared in Fig. \ref{fig:s_d_t}. The agreement between our MC and experimental data improves with increasing kinetic energy. This behaviour is indeed expected as the MC approach assumes that electrons are classical point-like particles. Moreover, our simulated spectra do not include the surface plasmon contribution, whose importance decreases at higher electron beam kinetic energies.    

\begin{figure}
    \centering
    \includegraphics[width=0.80\linewidth]{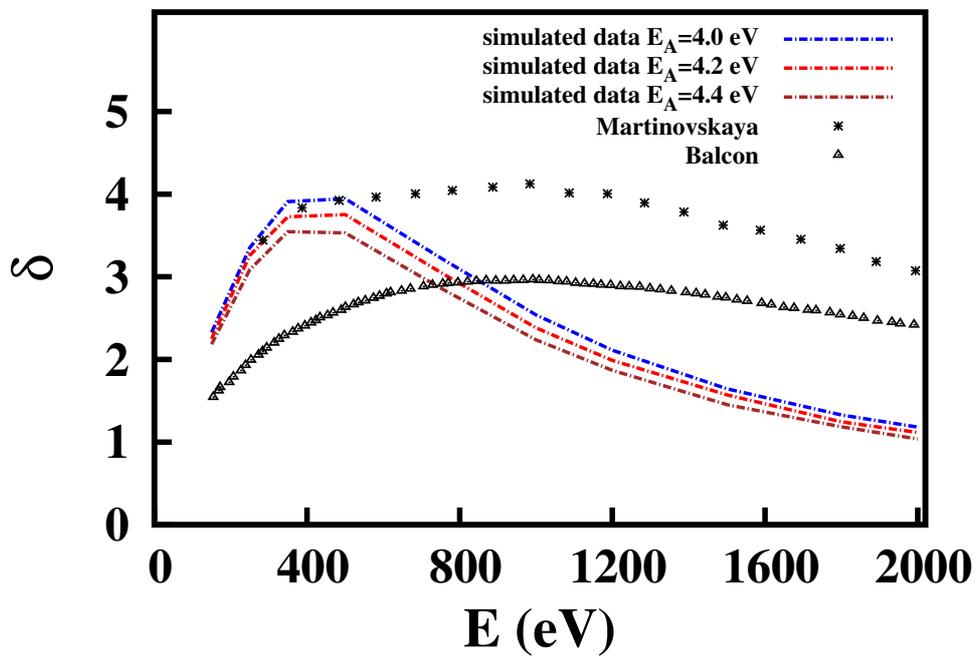}
    \caption{Monte Carlo simulated secondary electron yield $\delta$ of diamond as a function of the electron beam initial kinetic energy (eV) for several values of the electron affinity ($E_A$), using the full AI approach.}
    \label{fig:Dyield}
\end{figure}

Further Monte Carlo simulations were carried in order to obtain the secondary electron yield $\delta$. The inelastic scattering cross section was derived using the ELF obtained within the AI approach. Different values of electron affinity $E_A$ were considered in the range 1 to 5 eV and the simulations were carried out for several beam kinetic energies.
The results of our simulations are compared with the available experimental data \citep{NASA,neves2001properties} in Fig. \ref{fig:Dyield}.

The curve of the secondary electron yield shows a maximum at around 500 eV, while at lower kinetic energy electrons are 
easily trapped in the solid as they do not have enough kinetic energy to be ejected from the surface. At much higher initial kinetic 
energies electrons penetrate deeply into the solid where they undergo a large number of collisions and slow down to the point that they 
are not enough energetic to overcome the surface potential barrier.

\subsection{Graphite energy loss and secondary emission spectra}

\begin{figure}[ht!]
\centering
\includegraphics[width=0.98\linewidth]{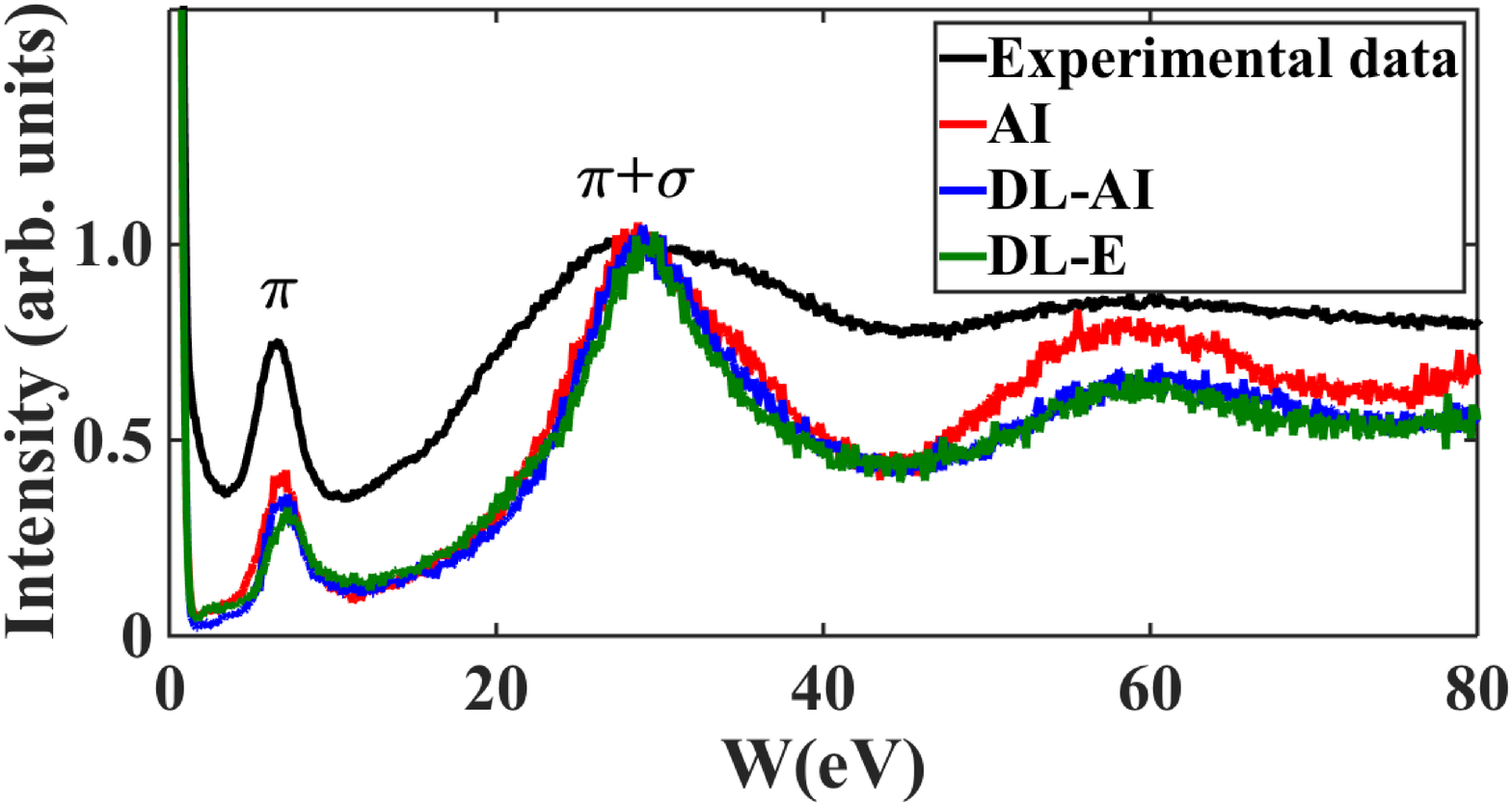}
\caption{REEL spectra of graphite: experimental data are reported in black, while simulations using the 3 different models are sketched in red (AI), blue (DL-AI) and green (DL-E). Electron beam kinetic energy is $1000$ eV. Data are normalized with respect to the $\pi+\sigma$ plasmon peak.}
\label{fig:prim_g}
\end{figure}

Graphite crystals were considered to have a density of 2.25 g/cm$^3$ \citep{Molina_NIMB_2006}. The band gap was set equal to $0.6$ eV according to our DFT calculations. 
MC simulations of REELS were carried out using the three different approaches to the calculation of the ELF mentioned above, with a number of electrons in the beam equal to $10^9$. 
Only the in-plane component of the energy loss function was dealt with in the calculation (i.e. we considered the component of the momentum transfer only along the single graphite layers). 
In Fig. \ref{fig:prim_g} we report the MC REELS simulations compared to our experimental measurements (black line).  

\begin{figure}[ht!]
\centering
\includegraphics[width=0.7\linewidth]{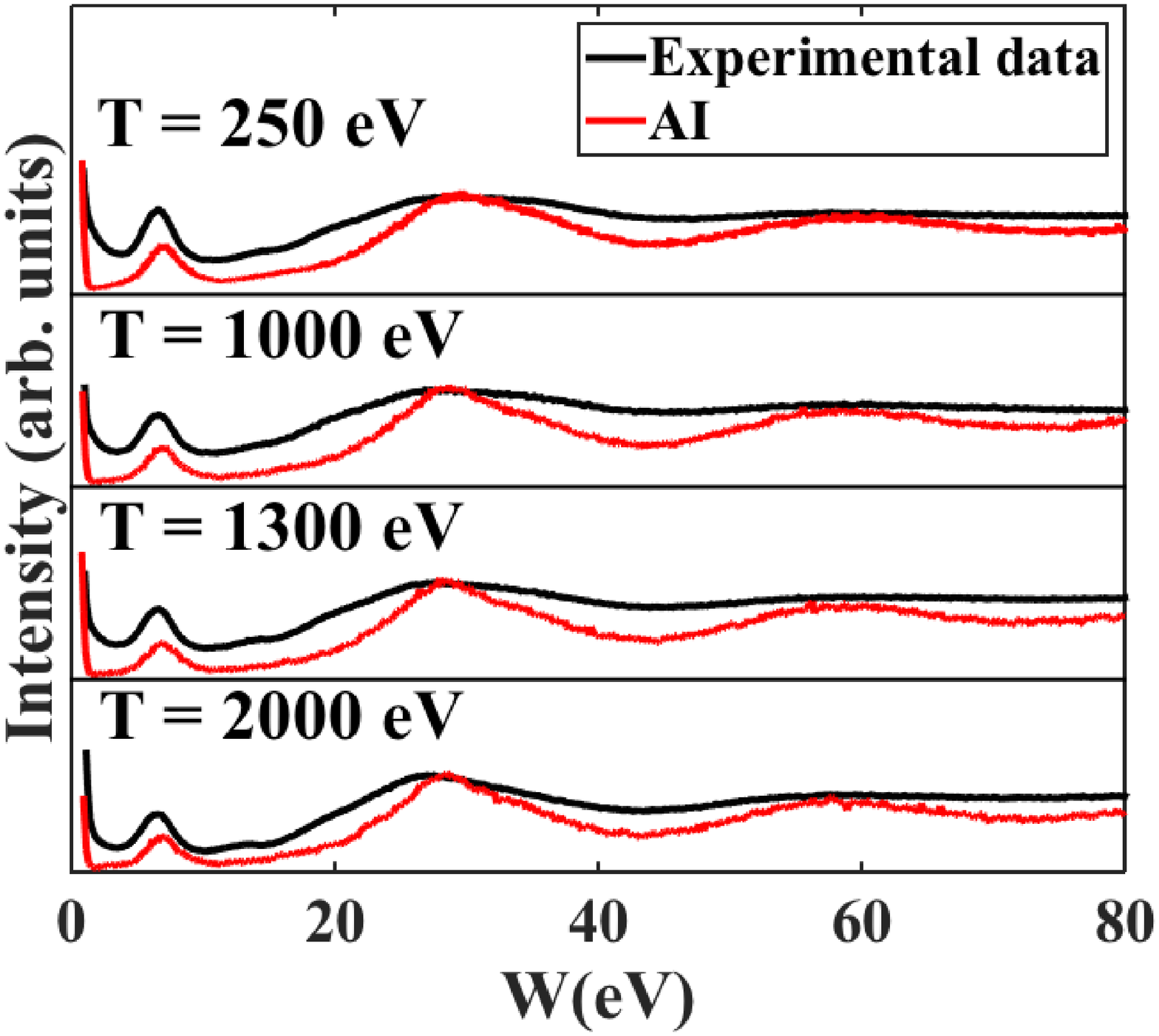}
\caption{REEL spectra of graphite obtained by calculating the ELF with the full AI approach (red line) at different momentum transfer compared to experimental data (black line) at different primary electron beam kinetic energy $T$. Data are normalized with respect to the $\pi+\sigma$ plasmon peak.}
\label{fig:prim_t_g}
\end{figure}

We notice that our MC simulations reproduce both the $\pi$ (due to the collective excitation of valence electrons in the $\pi$ band) and the $\pi + \sigma$ (due to collective excitation of all valence electrons) plasmon peaks. These findings are in agreement with the ELF function of Fig. \ref{fig:graph_back}), showing maxima at about same energies.
While the results of the simulations show a good agreement with experimental data independently of the ELF model, nevertheless, using the ab initio calculated ELF at finite momentum transfer, a third peak around 60 eV can be found. This peak corresponds to the two-plasmon excitation, and its presence is less apparent by adopting DL models. 
 Indeed in DL models the RPA approximation describes the system as composed by free electrons; in the case of graphite the electrons populating the $\pi$ bands are delocalized and they behave as almost-free electrons. For this reason the $\pi$ plasmon peak is well present in all three spectra. The discrepancies found in the energy loss spectral features advice the use of AI approaches for the extension of the ELF out of the optical region, in order to take into account accurately of the electronic motion inside the material. \\
REEL experimental spectra and MC simulated spectra obtained using the full $AI$ method are compared in Fig. \ref{fig:prim_t_g}. The agreement between our MC and experimental data improves with increasing kinetic energy, as for diamond.
  
\begin{figure}
    \centering
    \includegraphics[width=0.80\linewidth]{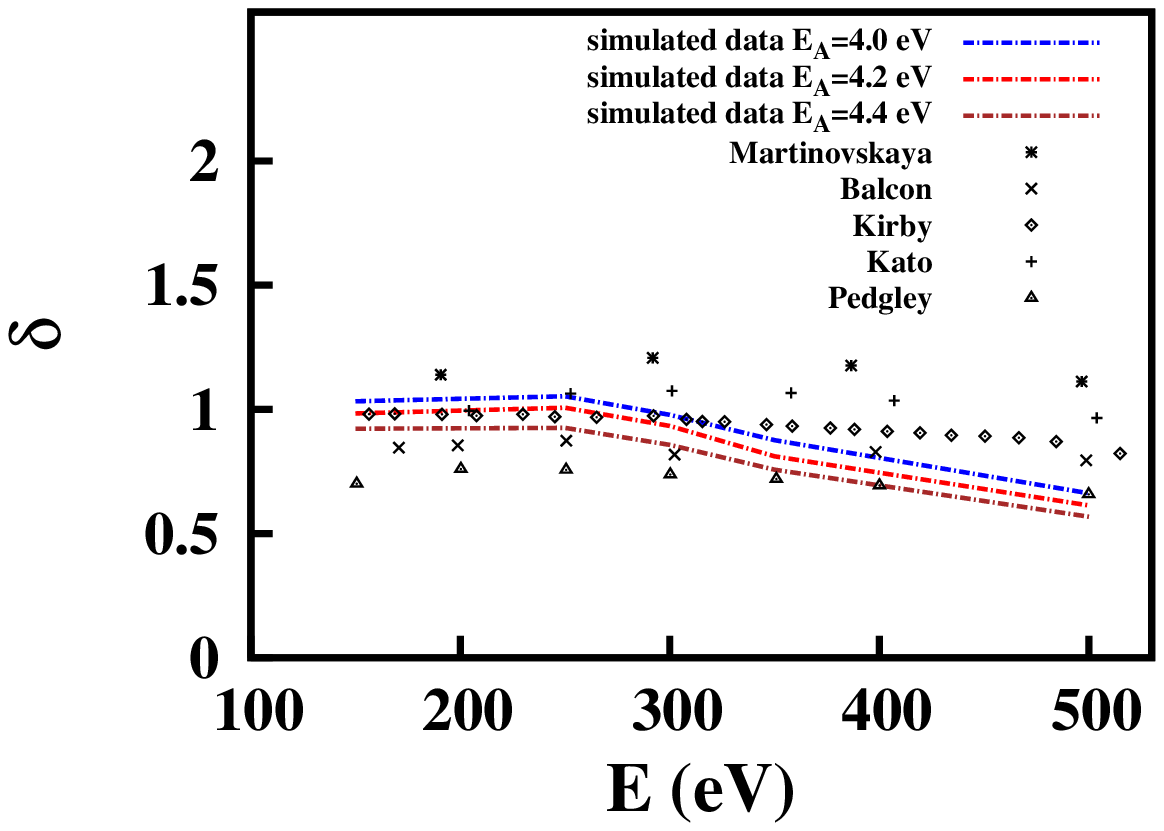}
     \caption{Monte Carlo simulated secondary electron yield $\delta$ of graphite as a function of the electron beam initial kinetic energy (eV) for several values of the electron affinity ($E_A$), using the full AI approach. Simulated data are compared with the available experimental data (Martinovskaya \citep{yieldG}, Balcon \citep{Balcon}, Kirby \citep{Kirby}, Kato \citep{Kato}, Pedgley \citep{Pedgley}).}
    \label{fig:Gyield}
\end{figure}

Using the full AI method for calculating the ELF, we plot in Fig. \ref{fig:Gyield} the secondary electron yield $\delta$ compared with available experimental data. The yield was calculated for several initial kinetic energy and by setting the value of the electron affinity in the range 3.8 eV to 4.6 eV. 

The yield displays a maximum at around $250$ eV, and shows a behaviour similar to diamond at all kinetic energies, where electrons are easily trapped in the solid.

\section{Conclusions}

In this work, we calculate the dielectric properties of diamond and graphite, in particular the dielectric response and the energy loss functions, using different approaches,
from semiclassical Drude--Lorentz to time-dependent density functional theory. We compare our computer simulations with REEL spectra recorded at several energies. 
The major result of this work is to point out that an accurate treatment of the electron-electron correlations beyond the random phase approximation of the homogeneous Fermi gas 
increases the overall accuracy of the model and provides a better agreement with measurements of back-scattered electrons.
This issue is particularly important in materials with highly dishomogeneous charge densities, such as semiconductors and insulators, 
out of the optical region where collective plasmon excitations mix with single particle-hole excitations. Semi-classical models, based on the Drude-Lorentz extension to finite momentum transfer, 
are generally less accurate in reproducing these quantum effects. Nevertheless, this finding turns out to be correct also for semimetals, such as graphite, even though in this case the accuracy of the
free-electron model (as the RPA) with respect to the energy loss spectral observables does not differ significantly from the ab initio model.
Thus, the advantage of using a full ab initio description of the electron scattering within the solid, particularly at low energy where the Monte Carlo approach is not rigorously applicable, clearly emerges by our computational analysis.\\ 
Finally, further developments of this work will be sought for the inclusion of the electron-phonon interaction, which is relevant at low energies, surface plasmons and anisotropy effects in 2D carbon materials. 

\section*{Acknowledgments}
N.M.P. is supported by the European Research Council (ERC StG Ideas 2011 BIHSNAM n. 279985, ERC PoC 2015 SILKENE nr. 693670) 
and by the European Commission under the Graphene Flagship (WP14 ``Polymer composites'', no. 696656). M.D., G.G., and S.T acknowledge 
funding from this last grant. This work used the ARCHER UK National Supercomputing Service (http://www.archer.ac.uk) and the FBK-KORE computing 
facility.

\end{document}